\documentclass[a4paper,fleqn,useAMS,usenatbib]{mnras}

\usepackage{graphicx,amsmath,amssymb,multirow,units,lpic}
\usepackage{lscape}
\usepackage[autolanguage]{numprint}
\usepackage{xcolor}
\usepackage{caption}
\usepackage[T1]{fontenc}
\usepackage{ae,aecompl}

\definecolor{purple}{rgb}{0.75,0.0,0.75}

\newcommand{\msun}{\mbox{\,$\rm M_{\odot}$\,}}        
\newcommand{\lsun}{\mbox{\,$\rm L_{\odot}$\,}}        
\newcommand{\lsub}{\mbox{\,$ L_{850}$\,}} 
\newcommand{\lcoa}{\mbox{\,$ L^{\prime}_{\rm{CO}}$\,}}

\newcommand{\Md}{\mbox{\,$M_{\rm d}$\,}}
\newcommand{\Ms}{\mbox{\,$M_{\ast}$\,}}

\newcommand{\Lir}{\mbox{\,$\rm L_{IR}$\,}}

\newcommand{\CI}{\,[C{\sc i}]\,}
\newcommand{\CIfull}{\mbox{\,[C{\sc i}]($^3$P$_1$-$^3$P$_0$)\,}}

\newcommand{\asec}{\ensuremath{^{\prime\prime}}\,}

\newcommand{\kms}{\mbox{\,$\rm{km\,s^{-1}}$\,}}
\newcommand{\td}{\mbox{\,$\rm{T_d}$\,}}

\newcommand{\asub}{\mbox{\,$\alpha_{\rm{850}}$\,}}

\newcommand{\mic}{$\mu $m\,}

\title[Too much of a good thing?]{Over-density of SMGs in fields containing $z\sim 0.3$
  galaxies: magnification bias and the implications for studies of galaxy evolution.}

\author[L. Dunne et al.]{
L. Dunne,$^{1,2}$ \thanks{E-mail:DunneL6@cardiff.ac.uk} 
L. Bonavera,$^{3,4}$
J. Gonzalez-Nuevo,$^{3,4}$
S. J. Maddox,$^{1}$
\and
C. Vlahakis$^{5}$
\\
$^{1}$School of Physics \&\ Astronomy, Cardiff University, Queens Buildings, The Parade, Cardiff, CF24 3AA, UK \\
$^{2}$SUPA, Institute for Astronomy, University of Edinbugh, Royal Observatory, Blackford Hill, Edinbugh EH9 3HJ, UK\\
$^{3}$Departamento de Fisica, Universidad de Oviedo C/ Federico Garcia Lorca, 18, 33007 Oviedo, Spain\\
$^{4}$Instituto Universitario de Ciencias y Tecnologias Espaciales de Asturias (ICTEA), C. Independencia 13, 33004 Oviedo, Spain\\
$^{5}$National Radio Astronomy Observatory, 520 Edgemont Road, Charlottesville, VA 22903-2475, USA\\
}

\date{\vspace*{-5em}\today}

\pubyear{2016}

\begin{document}
\label{firstpage}
\pagerange{\pageref{firstpage}--\pageref{lastpage}} 
\maketitle

\begin{abstract}
We report a remarkable over-density of high-redshift submillimeter
galaxies (SMG), 4--7 times the background, around a statistically
complete sample of twelve 250\mic selected galaxies at $z=0.35$, which
were targeted by ALMA in a study of gas tracers. This over-density is
consistent with the effect of lensing by the halos hosting the target
$z=0.35$ galaxies. The angular cross-correlation in this sample is
consistent with statistical measures of this effect made using larger
sub-mm samples. The magnitude of the over-density as a function of
radial separation is consistent with intermediate scale lensing by
halos of order $7\times 10^{13}\msun$, which should host one or
possibly two bright galaxies and several smaller satellites. This is
supported by observational evidence of interaction with satellites in
four out of the six fields with SMG, and membership of a spectroscopically defined
group for a fifth. We also investigate the impact of these SMG
on the reported {\em Herschel} fluxes of the $z=0.35$ galaxies, as they
produce significant contamination in the 350 and 500\mic {\em Herschel} bands. The
higher than random incidence of these boosting events implies a
significantly larger bias in the sub-mm colours of {\em Herschel}
sources associated with $z<0.7$ galaxies than has previously been assumed, with $f_{\rm boost} =
1.13, 1.26, 1.44$ at 250, 350 and 500\mic. This could have
implications for studies of spectral energy distributions, source
counts and luminosity functions based on {\em Herschel} samples at
$z=0.2-0.7$.

\end{abstract}


\section{Introduction}

The effect of gravitational lensing on our view of the distant
submillimeter (sub-mm) sky has been known and exploited since the
beginning of the era of sub-mm surveys. Lensing was originally seen as
a tool for gaining information on the fainter end of the source
counts, and surveys were targeted at massive clusters to benefit from
the strong magnifications they produce
\citep{SIB1997,blain99,cowie02,knudsen08,zemcov10,hsu16}. In more
recent years, strong lensing due to individual galaxies has also been
exploited to reveal the high redshift dusty star-bursts in exquisite
detail, with wide area FIR/sub-mm surveys producing large samples of such
strongly lensed systems
\citep[e.g.][]{Negrello2010,vieira10,canameras15,vlahakis15,swinbank15,negrello17,zhang18,jarugula19,yang19}.

The reason that lensing has been so helpful to sub-mm astronomy is
that the conditions for producing lots of lensing signal in the
population are optimal, with steep sub-mm source counts and the
population of SMG predominantly residing at high redshifts.
\citep[e.g.][]{blain97,negrello07,lapi12}. However, this boon from the
lensing phenomena comes with a price for those simply interested in
the statistics of the galaxy populations. Lensing is so ubiquitous
that it must be considered a possibility that any SMG which has a
luminosity $\Lir>10^{13}\lsun$ is being lensed, even if the lens
itself is not visible -- either being too high redshift, or because
the lensing is from intervening large scale structure such as a group
or cluster
\citep[e.g.][]{rowanrobinson91,graham95,Harris2012,bussmann15,nayyeri17}.

Statistical analyses comparing the positions of high redshift sub-mm
galaxies and lower redshift populations reported cross-correlation
signals right from the earliest days, but with only marginal
statistical significance given the small areas imaged
\citep{almaini05,aretxaga11,wang11}. With the wide area submillimeter
      {\em Herschel}-ATLAS survey \citep{Eales2010} and complementary
      optical spectroscopic survey from GAMA \citep{Driver2011}, a leap
      was made in the detection and characterisation of this signal as
      being that of cosmological lensing bias: the lensing effect from
      the foreground large-scale structure on the background high
      redshift galaxy population \citep{GN14,GN17}. The lensing is
      not thought to be strong, with magnification factors of 1.0--1.5
      but it nevertheless changes the statistics and potentially imprints the
      correlation function of the high-z galaxies with the signal from
      the lower redshift structures which are magnifying them.

The two regimes of lensing, strong (from clusters
  or single galaxies producing arcs, rings or multiple images), and
  weak (from large scale foreground structure which produces the
  cosmological lensing bias at large angular scales seen in
  \citet{hildebrandt13} and \citet{GN14,GN17}), are reasonably well
recognised and understood. There is, however, a third regime
intermediate between strong and weak which has only recently been
identified, and which is the subject of this paper. This regime,
spanning scales of a few arcsec to a few tens of arcsec, produces an
upturn in the cross-correlation signal between low redshift
populations and $z>1.5$ SMG. Statistical evidence
  for this intermediate lensing regime has been highlighted in the
  angular cross-correlation study of H-ATLAS high redshift sources
  with low redshift optical galaxies \citet{GN17}. Pre-dating this
study, it was also hypothesised as a possible explanation for a
puzzling trend noted during the H-ATLAS cross-identification studies
\citep{DJBSmith2011,Bourne2016} in which H-ATLAS sources with red sub-mm
colours (aka high redshift) and SDSS optical galaxies had a broader
cross-correlation peak in angular scale compared to H-ATLAS sources
with blue sub-mm colours (aka low redshift) and the same SDSS optical
galaxy catalogue \citep{Bourne2014}. Both statistical signals are
thought to be due to the same effect: moderate lensing by halos
hosting galaxy groups or very massive centrals with a number of
satellite dwarfs. The lensing is not strong enough to create
distortions in the sub-mm images, but should have amplifications in
the range $\mu=1-3$ \citep{GN17}. The lensing occurs on scales similar
to the profile of the large halo or group of halos, so is at 3-15\asec
rather than the $<3\asec$ expected for strong galaxy-galaxy lensing,
or the $>1\arcmin$ scales of weak lensing by the large scale
structure.
 
In this paper we describe the serendipitous detection of a large
over-density of SMG within 13\asec of $z=0.35$ galaxies. The galaxies were
originally targeted by ALMA as a small, but homogeneously
selected sample in the relatively local Universe, which could provide
a calibration of gas tracers in dust, CO and \CI. We believe that this
is a first detection of the intermediate lensing regime in individual sources. 

In Section~\ref{Sample} we describe the sample,
observations and data reduction. In Section~\ref{smgS} we present the
very surprising result that there is an over-density of a factor 4--6
in the number of SMG found in these fields compared to blank field
surveys. In Section~\ref{lensS} we investigate a lensing mechanism for
this over-density and in Section~\ref{boostS} we discuss the wider
implications of flux boosting in {\em Herschel} surveys.  Throughout
we use a cosmology with $\Omega_m = 0.308,\,\Omega_{\Lambda} =0.692$ and
$H_o = 67.8\, \rm{km\,s^{-1}\,Mpc^{-1}}$ \citep{planckcosmo}.

\section{Sample and Data}
\label{Sample}

\begin{table*}
\caption{\label{sampleT} Properties of the target $z=0.35$ galaxies.}  \centering
\begin{tabular}{lcccccccc}
\hline
H-ATLAS IAU & SDP & GAMA  & R.A & Dec. & z & $S_{250}$ & $\rm{r_{pet}}$   & SMG \\  
            & name & ID   & hh:mm:ss.s & dd:mm:ss.s   &   & (mJy)    & (mag) &     \\
\hline
J$090506.2+020700$ & 163  & 347099 & 09:05:06.1 & 02:07:02.2 & 0.345 & 107.6 & 18.8 &   N \\
J$090030.0+012200$ & 1160 & 301774 & 09:00:30.1 & 01:22:00.2 & 0.353 & 48.4  & 19.15 &  Y\\
J$085849.3+012742$ & 2173 & 376723 & 08:58:49.4 & 01:27:41.0 & 0.355 & 46.2  & 18.71 &  Y\\
J$091435.3-000936$ & 3132 & 575168 & 09:14:35.3 & $-$00:09:35.6 & 0.359 & 40.6 & 19.02&  N\\
J$090450.0-001200$ & 3366 & 574555 & 09:04:50.1 & $-$00:12:03.0 & 0.354 & 40.3 & 18.93&  Y\\
J$090707.7+000003$ & 4104 & 210168 & 09:07:07.9 & 00:00:02.1 & 0.350 & 46.2 & 19.38 &  Y\\
J$090845.3+025322$ & 5323 & 518630 & 09:08:45.3 & 02:53:20.0 & 0.353 & 28.6 & 18.98 &    N\\
J$090658.6+020242$ & 5347 & 382441 & 09:06:58.4 & 02:02:44.7 & 0.347 & 32.7 & 19.01 &  Y\\
J$090444.9+002042$ & 5526 & 600545 & 09:04:44.9 & 00:20:48.2 & 0.342 & 31.2 & 19.23 &  N\\
J$090844.8-002119$ & 6216 & 204249 & 09:08:44.8 & $-$00:21:18.0 & 0.352 & 36.2 & 18.75  & N \\
J$090402.3+010800$ & 6418 & 372500 & 09:04:02.2 & 01:07:58.2 & 0.347 & 31.6 & 18.96 &  Y\\
J$090849.4+022557$ & 6451 & 387660 & 09:08:49.5 & 02:25:56.9 & 0.353 & 33.7 & 19.08 &   N\\
\hline
\end{tabular}
\flushleft{\small{\bf Notes:} Positions and redshifts refer to the
  optical properties of the XID in \citet{Bourne2016}. 250\mic flux is
  from the H-ATLAS DR1 release \citep{Valiante2016}. $\rm{r_{pet}}$ is
  the SDSS r-band petrosian magnitude from GAMA. SMG indicates whether
  a background high-redshift SMG is detected at $>5\sigma$ in the Band
  7 field.}
\end{table*}

\begin{table*}
\caption{\label{photomT} FIR photometry for the $z=0.35$ galaxies from {\rm
    Herschel-ATLAS} DR1 and our ALMA measurements.}
\begin{tabular}{lcccccccccccc}
\hline
Source & $S_{100}$ & $\sigma_{100}$ & $S_{160}$ & $\sigma_{160}$ & $S_{250}$ & $\sigma_{250}$ & $S_{350}$ & $\sigma_{350}$ & $S_{500}$ & $\sigma_{500}$ & $S_{850}$ & $\sigma_{850}$\\
\hline
163 & 73.9 & 20.7 & 102.1  & 24.1 & 107.6 & 7.3 & 50.7 & 8.1 & 23.9 & 8.5 & 3.05 & 0.34\\
1160$^\dag$ & 49.8 & 17.5 & 57.1 & 19.7 & 48.4 & 7.2 & 32.4 & 8.1 & 21.6 & 8.7 & 0.54 & 0.12 \\
2173$^\dag$ & 59.0$^M$ & 34.0 & 68.0$^M$ & 38.0 & 46.2 & 6.5 & 21.4 & 7.5 & 11.1 & 7.8 & 0.74 & 0.20\\
3132 & 64.3$^M$ & 26.5 & 65.1$^M$ & 20.0 & 40.6 & 6.4 & 23.6 & 7.4 & 13.1 & 7.8 & 0.95 & 0.29\\
3366$^\dag$ & 19.7 & 24.5 & 114 & 37.6 & 40.3 & 7.3 & 26.3 & 8.0 & 16.5 & 8.8 & 0.42 & 0.019\\
4104$^\dag$ & 77.9 & 17.6 & 53.3 & 26.4 & 46.2 & 7.2 & 28.3 & 8.1 & 12.1 & 8.8 & 0.92 & 0.24\\
5323 &  ..  & ..   & ..   & ..   & 28.6 & 7.1 & 30.0 & 8.0 & 9.6 & 8.4 & 0.89 & 0.24\\
5347$^\dag$ & 33.0 & 41.2 & 68.0 & 17.7 & 32.7 & 7.5 & 29.8 & 8.2 & 17.2 & 8.7 & 1.84 & 0.35\\
5526 & 62.0$^M$ & 32.6 & 56.9$^M$ & 40.7 & 31.2 & 7.3 & 20.0 & 8.2 & $-$10.7 & 8.5 & 1.11 & 0.27\\
6216 & 35.0$^M$ & 36.6 & 34.0$^M$ & 40.7 & 36.2 & 7.3 & 19.9 & 8.1 & 3.8 & 8.8 & 1.49 & 0.25\\
6418$^\dag$ & 40.4$^M$ & 19.8 & 27.0$^M$ & 33.3 & 31.6 & 7.3 & 18.5 & 8.0 & 16.6 & 8.4 & 1.06 & 0.32\\
6451$^\dag$ & 69.4 & 41.7 & 59.5 & 47.8 & 33.7 & 7.3 & 29.7 & 8.2 & 19.5 & 8.6 & 0.88 & 0.19\\
\hline
\end{tabular}
\flushleft{\small{\bf Notes:} Fluxes are all in mJy. $^{\dag}$
  indicates that there is evidence for contamination of these {\em
    Herschel} fluxes by high-z SMG in the beam. Before SED fitting we
  subtract from these fluxes the estimated contamination from high-z SMG listed in
  Table~\ref{SMGcontamT}. See S~\ref{boostS} for details. $^M$ indicates a
  PACS flux re-measured from the DR1 maps.  }
\end{table*}

The aim of the proposed ALMA observations was to map CO, dust and
\CIfull in a sample of 250\mic selected galaxies to make a calibration
of molecular gas mass based on three tracers. Full details of the
project and results from the dust, CO and \CI imaging are presented in
a companion paper, Dunne et al. {\em submitted}. The sample was selected
from the {\em Herschel}-ATLAS \citep{Eales2010} Science Demonstration
Phase (SDP) equatorial field at R.A. 09h. H-ATLAS is the {\em first
  unbiased survey of the dust content of local galaxies}, covering 660
sq. deg and sensitive to the cold dust component which dominates the
mass of dust in galaxies. It was the widest area extragalactic survey
carried out with the {\em Herschel} Space Observatory
\citep{Pilbratt2010}, imaging 600 deg$^2$ in five bands centred on
100, 160, 250, 350 and 500$\mu$m, using the PACS \citep{Poglitsch2010}
and SPIRE instruments \citep{Griffin2010}. The {\em Herschel}
observations consist of two scans in parallel mode reaching a
4$\sigma$ point source sensitivity of 28 mJy beam$^{-1}$ at
250\mic. The angular resolution is approximately 9\asec, 13\asec,
18\asec, 25\asec\ and 35\asec in each of the five bands. While the
original sample for the proposal was selected from the SDP public
release catalogue described in \citet{Rigby2011} to have
$S_{250}>5\sigma$ and a reliable optical identification with spectroscopic
redshift from \citet{DJBSmith2011}, we update the Herschel photometry
and optical parameters in this paper to those from the H-ATLAS DR1
release \citep{Valiante2016,Bourne2016}. 

In order to fulfil the requirements of the ALMA Cycle 1 call where
only Band 7 and Band 3 were available, all the sources had to be
within 12 degrees of each other on the sky and had to be observed using no
more than five tunings, resulting in a very limited redshift range
around $z=0.34-0.36$ that fulfilled these requirements. We selected
{\em all} the H-ATLAS SDP sources within this redshift range, making
this sample of twelve representative of sources from a blind 250\mic
selected sample at $z\sim 0.3-0.4$. Details of the sample are given in
Table~\ref{sampleT}.

The H-ATLAS DR1 photometry we use is given in Table~\ref{photomT}, to
which we add the ALMA 850\mic fluxes for the $z=0.35$ sources which
are taken from Dunne et al. {\em submitted}. We use the SPIRE matched
filter photometry from the DR1 release, as these are all point sources
and this is the most likely estimate of their flux \citep{MADX}. In
the case of PACS, the LAMBDAR algorithm of \citet{Wright2016}
produces, in our opinion, a more robust measure of the PACS fluxes and
errors as instead of using a top hat aperture, it convolves the
optical r-band aperture with the PACS PSF and so measures flux in a
PSF-weighted aperture. However, in 5 cases there was a significant
difference between the two catalogues, so we returned to the original
H-ATLAS PACS maps and remeasured our own photometry (indicated with
$^M$ in Table~\ref{photomT}). Spectroscopic redshifts and UV-22\mic
photometry are provided by the Galaxy and Mass Assembly (GAMA) survey
\citep{Driver2011,Liske2015,Wright2016}.

\begin{figure}
\includegraphics[width=0.49\textwidth]{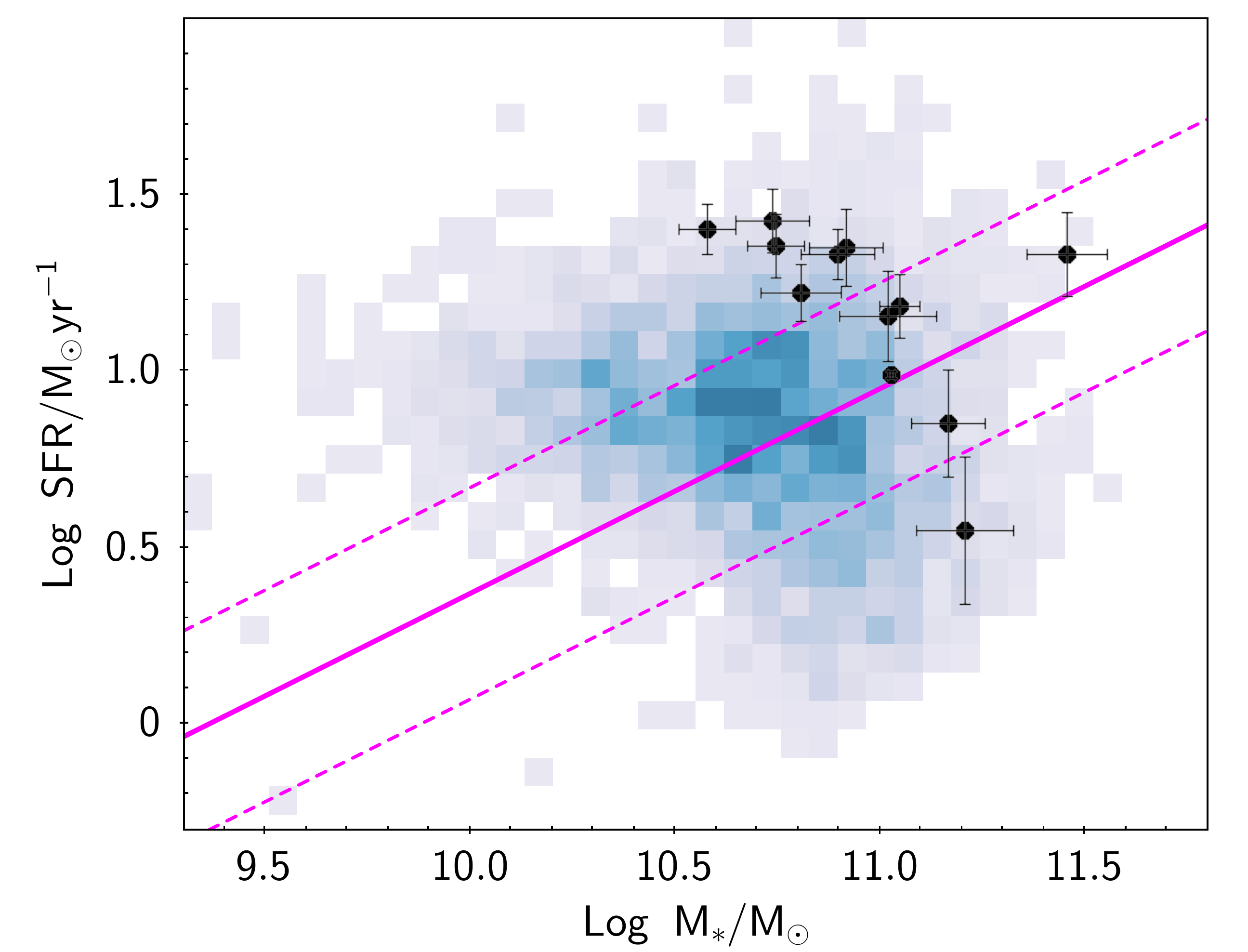}
\caption{\label{MSplotF} Star Formation Rate (SFR) versus stellar mass for the $z=0.35$
  sources with the best fit Main Sequence (solid) and $\pm0.3$ dex
  intrinsic scatter (dashed lines) from \citet{Speagle2014}. The
  coloured region represents the density of galaxies in the range
  $0.33<z<0.36$ with log $\rm{sSFR>-11.0}$ and
  $\rm{\sigma_{M_{\ast}}<0.15,\,\sigma_{SFR}<0.2}$ from GAMA
  \citep{Driver2016}. This also shows the effect of the 250\mic
  selection at this redshift, where we sample the leading edge of the
  distribution in SFR-\Ms for the optical selection $r<19.8$ from
  GAMA.}
\end{figure}

The sample has a narrow range of $\rm{L_{IR}} =
1.2\times10^{11}-6\times10^{11}\,\lsun$, making them far more `typical'
of galaxies at this redshift than previous very luminous IR samples
\citep[e.g.][]{Combes2011}. The stellar masses are in the range
$\Ms=4\times10^{10}-3\times10^{11}\,\msun$. Comparison of the 250\mic
selected sources with other optically selected galaxies at the same
redshift from the GAMA survey \citep{Driver2016, Baldry2018} in
Figure~\ref{MSplotF} shows that the 250\mic selection picks out the
leading edge of the optical cloud of galaxies, i.e. only the most
massive or highly star forming galaxies at this redshift make it above
the {\em Herschel\/} flux limit.

\subsection{ALMA observations and data reduction}
\label{COdata}

\begin{table*}
\caption{\label{ALMAparamsT} Details of the ALMA observing for Band 7.}  \centering
\begin{tabular}{lccccccccccc}
\hline
SB name & $\rm{\nu_{obs}}$ &  Date & MS & $N_{\rm ant}$ & $t_{\rm int}$ & p.w.v &  Phase Cal. & Flux Cal. & B.P. &  $\sigma_{\rm{cont}}$ \\
        & (GHz)           &        &        &              &  (min)    & (mm)   &               &          &      &   ($\mu$Jy)\\
\hline
LowV   & 365.2  & 14.12.13  & X1644 & 26 & 50.2 & 0.84  & J$0854+20$ & Pallas  & J$1058+01$ & 90 &0.86 -- 1.2 \\ 
       &            & 26.12.14  & X2543 & 40 & 33.3 & 0.84  & J$0909+01$ & Ganymede & J$0825+03$ & .. &  \\ 
HighV      & 363.5 & 21.2.14 & X3e5 & 28  & 35.3 & 0.64 &J$0914+02$ & Ganymede & J$0825+03$  & 65 & 0.84 -- 1.1\\
           &       & 2.1.15 & X1e75 & 39  & 34.2 & 0.69 &J$0909+01$ & Callisto & J$0825+03$  & .. & \\
           &       & 2.1.15 & X2365 & 39  & 15.4 & 0.82 &J$0909+01$ & Ganymede & J$0825+03$  & ..&  \\
           &       & 2.1.15 & X2606 & 39  & 34.2 & 0.91 &J$0909+01$ & Ganymede & J$0909+01$  & ..&  \\
\hline
\end{tabular}
\flushleft{\small $\sigma_{\rm{cont}}$ is
  the average rms in $\mu$Jy/beam for the continuum.}
\end{table*}

Observations in the 3-mm band were made in Dec 2013 during Cycle 1
with the ALMA Band 3 receiver tuned to 85 GHz. The total integration
time for all 12 sources plus calibrations was 96 minutes giving
$\sigma_{\rm cont}= 40\mu$Jy per synthesised beam of
$\theta_3=2.4\asec \times1.8\asec$. The \CIfull observations were
split across Cycles 1 and 2 spanning a period from December
2013--January 2015, and using 4 tunings of the Band 7 receiver in the
range 362--367 GHz. The total integration time was 10.7 hours giving
$\sigma_{\rm cont}=65-90\mu$Jy beam$^{-1}$ with $\theta_7= 1.03\asec
\times0.64\asec$. A list of the observations is presented in
Table~\ref{ALMAparamsT}.

The 3-mm Band 3 data were reduced manually using the Common
Astronomy Software Applications (CASA) v4.5 package
\citep{McMullin2007} with flux calibration from Mars and the
phase/bandpass calibrator J0854+2006. The two measurement sets (MS),
which were observed on the same day, were concatenated before imaging
having been set to a common flux scale. We created spectral line cubes
using {\sc tclean} in CASA in 100\kms channels, natural
weighting was used in order to maximise signal-to-noise. We also
imaged in spectral line mode, the three TDM windows which were used
for the continuum at 3-mm. We did so in order to search for emission
lines in the high redshift SMG discovered in our fields. The noise in
these cubes was 0.4--0.5 mJy beam$^{-1}$ respectively.

The Band 7 observations were set up in four Scheduling Blocks (SBs)
where the sources that could share a single tuning were grouped into a
given SB. Due to some of these data being taking during Cycle 2, the
newer CASA v4.7 was used for the reduction. Some of the MS were
calibrated using the ALMA pipeline, while others were reduced
manually, depending on how the data were delivered. All MS were
checked and reprocessed allowing for tailoring of the calibration to
the specific issues in this data-set. For example, there are several
atmospheric lines which are evident in the Band 7 data and so the
pipeline calibration was modified to avoid flagging the system noise
temperature ($\rm{T_{sys}}$)\footnote{
    $\rm{T_{sys}}$ is a representation of the noise from both
    receivers and the atmosphere and is used to calculate initial
    weightings of the visibility data. In spectral regions of high
    atmospheric opacity there is a resultant spike in $\rm{T_{sys}}$
    with frequency, and as these spikes are genuine reductions in
    sensitivity in this part of the spectrum they should not be clipped
    from the $\rm{T_{\rm sys}}$ calibration curves which are used in
    future calibrations.} response in some cases, and to output the
data-set from the pipeline after the generation of the water vapour
radiometer (WVR) and $\rm{T_{sys}}$ calibration tables. The data were
then manually processed from that stage so that the atmospheric lines
could be flagged in the bandpass calibrator and the bandpass solutions
interpolated in these regions. Additional manual flagging was also
applied where required.

Imaging in B7 between 350-360~GHz was performed using the {\sc casa}
task {\sc tclean} with a Hogbom algorithm, using natural
weighting. Sources above $4\sigma$ in the dirty image were masked and
lightly cleaned (to 1.5$\sigma$).

\begin{table*}
\caption{\label{robustT} Robustness check of detection of SMG in the individual data-sets. }
\begin{tabular}{lcccccccc}
\hline
EB   & Typical rms & \multicolumn{7}{c}{Signal-to-noise ratio of the SMG in each individual EB.} \\
     & ($\mu$Jy beam$^{-1}$) & 1160.s1 & 1160.s2 & 2173 & 3366 & 4104 & 5437 & 6418 \\
\hline
X3e5  & 136 & 5.7    & 3.4  & 3.8 & 5.5  &  &  &  \\
X1e75 & 100 & 7.9    & 8.4  & 5.3 & 4.6  & &  &  \\
X2365 & 171 & 3.5    & 4.6  & 3.6 & 3.9  & &  &  \\
X2606 & 170 & 4.8    & 3.1  & 3.8 & 5.5  & &  &  \\
X1644 & 163 &        &                &           &                &  3.7 & 3.5 & 12.8\\
X2543 & 103 &        &                &           &                &  5.0 & 7.2 & 15.2\\
\hline
\end{tabular}
\flushleft{Notes: This table records the SNR of each SMG in all of the individual Execution Blocks. In every case, there is a positive source at the same position in each EB. Sources brighter than 4$\sigma$ were cleaned, there is no significant change in properties with or without cleaning applied (except for the brightest source 6418, which is bright enough to self calibrate).}
\end{table*}

\begin{table}
\caption{\label{negtestT} Results of inverting the images for the SMG fields}
\begin{tabular}{lccccc}
\hline
Field & \multicolumn{5}{c}{Number of peaks in given signal-to-noise range}\\
& $(<-5)$ & ($-5,-4$) & ($4,5$) & ($5,6$) & ($>6$) \\
\hline
1160  & 0          & 1          & 1$^t$       & 0        & 2\\
2173$^K$  & 0          & 0          & 1           & 0        & 1\\
3366$^K$  & 0          & 1          & 1           & 0        & 1\\
4104$^K$  & 0          & 0          & 1$^t$       & 1        & 0\\
5347$^K$  & 0          & 0          & 0           & 0        & 1\\
6418  & 0          & 0          & 0           & 0        & 1\\
\hline
\end{tabular}
\flushleft{Notes: Number of negative peaks more significant than
  $-4\sigma$ compared to the positive peaks as a simple test of source
  robustness. There are no negative peaks at $<-5\sigma$ while there
  are 7 peaks at $>+5\sigma$. A $^K$ symbol at the field name
  indicates that the SMG has a K-band counterpart. A $^t$ symbol
  indicates that this positive peak is the target low-z galaxy.}
\end{table}

During imaging, it was obvious from the dirty images that there were
point sources in the fields of at least half of the targets, which
were not associated with the $z=0.35$ system. They were very obvious
as they were usually much higher SNR than the target source itself. We
did not run any source extraction algorithms, but merely noted when a
bright point source was present in the imaging (as it needed to be
cleaned) and later on went back to analyse these serendipitous
detections. To check the robustness of these sources, we made two
analyses. Firstly, we noted the flux and SNR of each SMG in all of the
different Execution Blocks (EB) in which they were observed. Four of
the SMG were observed in four execution blocks, three on the same
night and another taken almost a year earlier. The other three SMG
were observed in two execution blocks, taken roughly one year
apart. Table~\ref{robustT} shows the SNR ratios for each SMG in each
of the EB imaged separately. In every case, the SMG is present in each
of the EB going into the final data-set. Obviously, those EB with
shorter integration times, poorer weather or fewer antennae have lower
SNR for the SMG, but nevertheless they are positive peaks above
3$\sigma$ in each of the separate observations. The resulting sample
of seven robust SMG are all those detected with SNR$>5\sigma$, (in
fact all of them have SNR$\geq 6\sigma$). There are also two peaks
just below the 5$\sigma$ threshold, which are listed as candidate
sources in Table~\ref{candidateT} but are not considered further in
the analysis. The second test that we did was to count the negative
and positive peaks in the final images. The results are shown in
Table~\ref{negtestT}, and in no cases are any negative peaks at or
below $-5\sigma$ found. Furthermore, four of the SMG (with the lowest
SNR) also are found to have K-band counterparts in infrared imaging
from the VIKING survey \citep{Edge2013}. This combined with the
detection of all SMG in all the individual EB confirms that this is a
very robust set of sources. Given the number of beams in the surveyed
area, we expect a false detection rate at 5$\sigma$ of 0.002 sources
in this sample.

\subsection{Flux and size measurements}

\begin{figure*}
\includegraphics[width=0.45\textwidth,trim=0.0cm 0.0cm 0.0cm 0.0cm, clip=true]{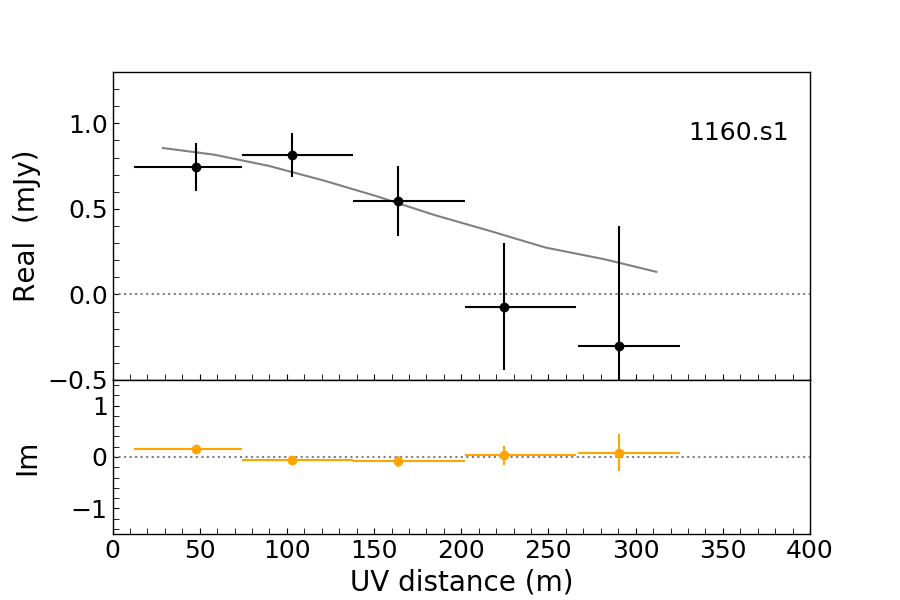}
\includegraphics[width=0.45\textwidth,trim=0.0cm 0.0cm 0.0cm 0.0cm, clip=true]{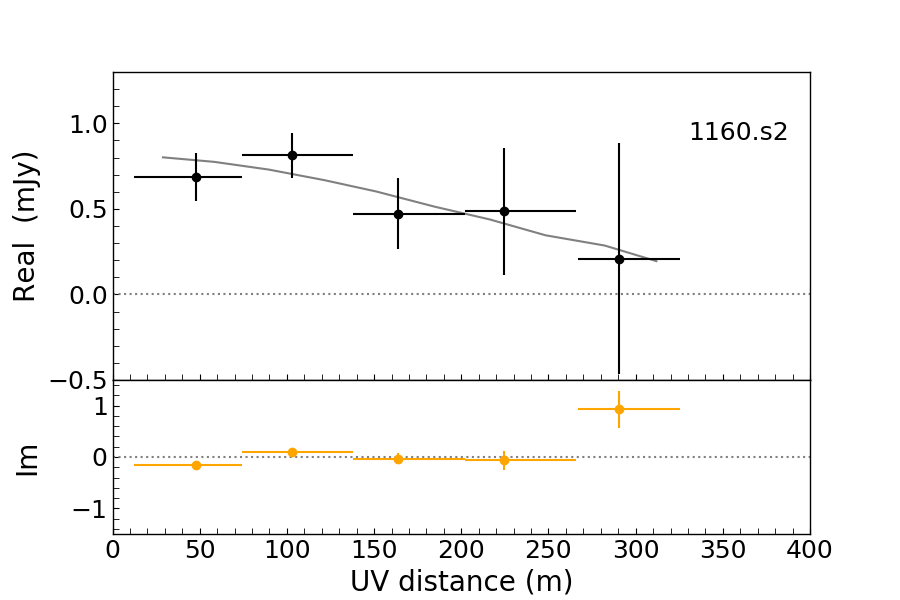}\\
\includegraphics[width=0.45\textwidth,trim=0.0cm 0.0cm 0.0cm 0.0cm, clip=true]{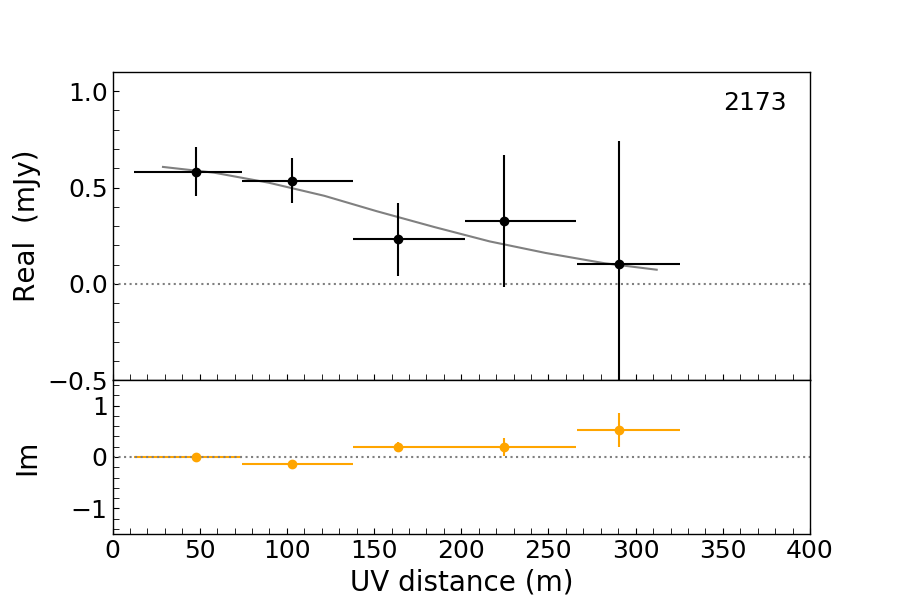}
\includegraphics[width=0.45\textwidth,trim=0.0cm 0.0cm 0.0cm 0.0cm, clip=true]{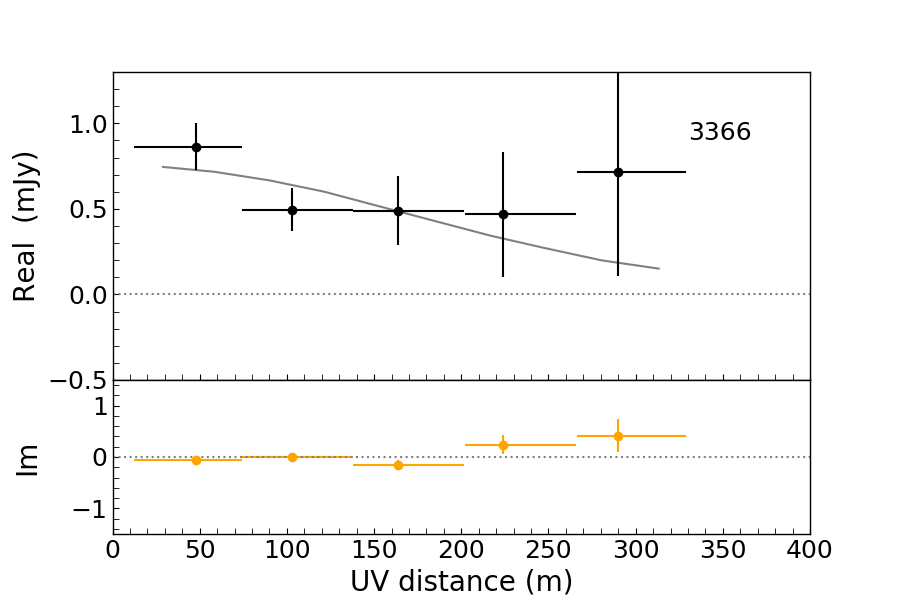}\\
\includegraphics[width=0.45\textwidth,trim=0.0cm 0.0cm 0.0cm 0.0cm, clip=true]{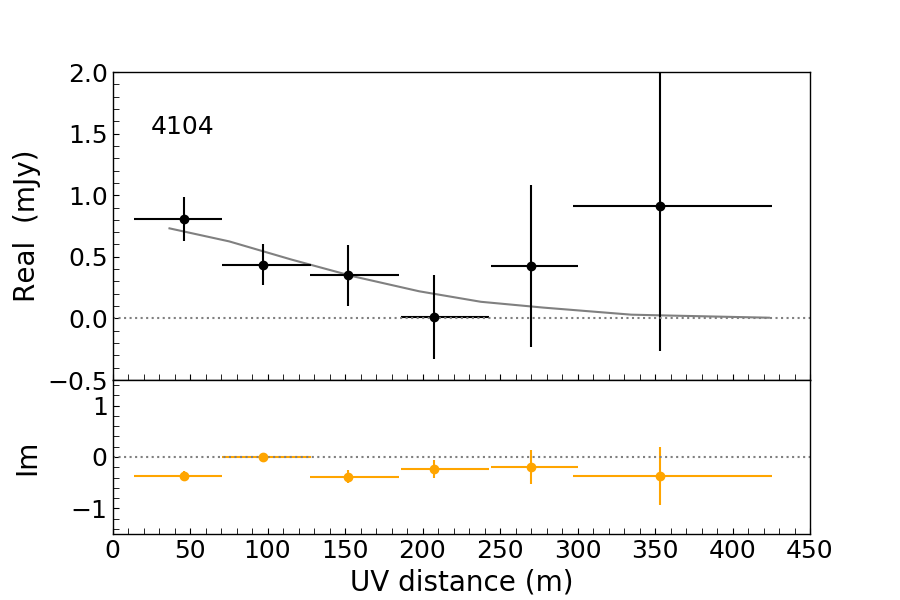}
\includegraphics[width=0.45\textwidth,trim=0.0cm 0.0cm 0.0cm 0.0cm, clip=true]{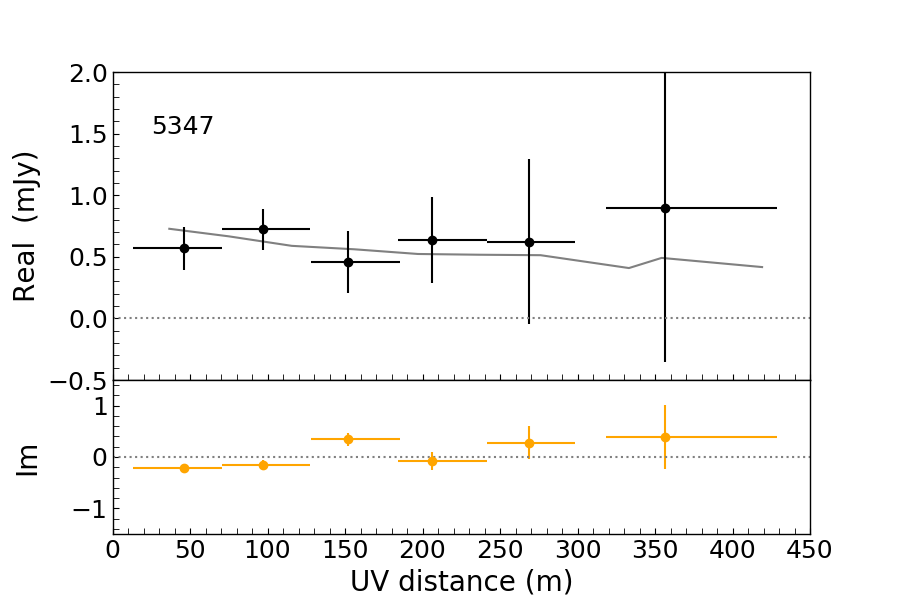}\\
\includegraphics[width=0.45\textwidth,trim=0.0cm 0.0cm 0.0cm 0.0cm, clip=true]{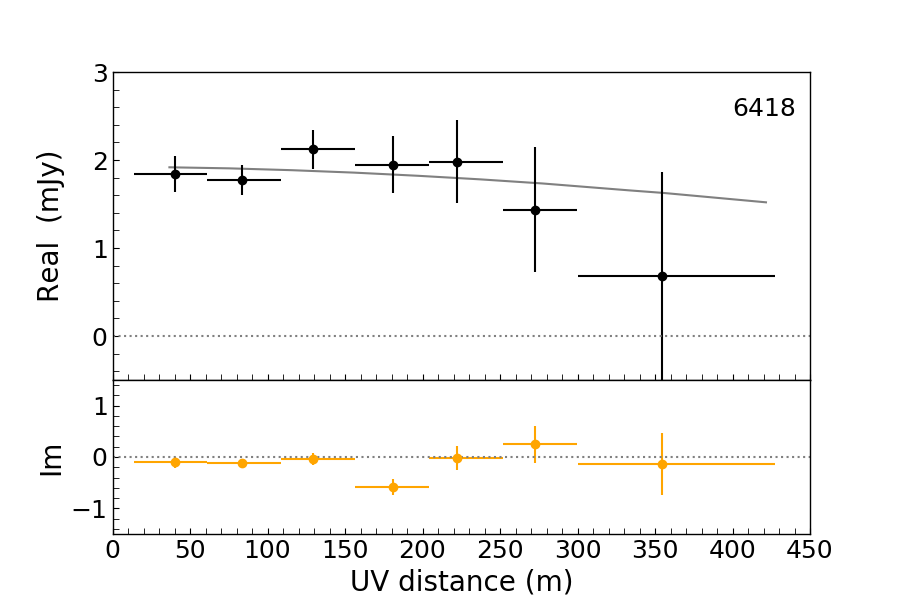}
\caption{\label{uvfitF} The binned $u,v$ data with Gaussian fits from
  {\sc uvmodelfit} over-plotted for the 7 SMG. Details of the fit
  parameters are in Table~\ref{robustT}. In all cases the data were
  averaged in frequency to one point per spectral window and in time
  by 30 seconds before fitting. Points are averaged in radial bins of
  u-v distance using the weights from {\sc CASA}. Error bars on points
  are estimated from the scatter of the data points within each bin
  (this is comparable but slightly larger than using the inverse
  square-root of the sum of the weights). }
\end{figure*}

We measured the fluxes of the source in two ways, firstly by fitting a
Gaussian and taking the peak flux from the CASA task {\sc imfit}, and
secondly, by fitting a Gaussian in the $(u,v)$ plane using {\sc
  uvmdodelfit} \citep{MartiVidal2014}. We used both methods as a check for systematics, and
also because very little size information can be derived in the image
plane for sources which are smaller than the synthesised beam. Fitting
in the $(u,v)$ plane is more direct and avoids the uncertainties
associated with the non-linear process of deconvolution in the
imaging. The fluxes and sizes are reported in Table~\ref{SMGT}, the
first row for each source lists the parameters from the $(u,v)$
fitting while the second row lists the peak flux reported by {\sc
  imfit} (equivalent to a point source flux).



For the $u,v$ fitting we averaged the data across each spectral window
and in intervals of 30 seconds, both of these do not lead to
significant band-width or time smearing for this ALMA band. We then
fixed the phase centre to be the position of the SMG and fitted with a
2D Gaussian where the flux $S_{\rm fit}$, position, major axis size
$\theta_{m1}$ and position angle ($pa$) were free parameters. The
sixth parameter, the axial ratio $b/a$, was found to be unconstrained
by fitting, tending to an axial ratio approaching zero, usually
aligned with the $pa$ of the beam. This tendency produces a bias in
the values for $S_{\rm fit}$ and $\theta_{m1}$, both of them being
maximised. Since an axial ratio approaching zero is not physical, we
have used a constrained fit as described in
Appendix~\ref{sec:inclination}.

The major axis source sizes fitted in this way range from
0.1--0.6\asec, and the average value of the circularised size is
$\theta = 0.36$\asec, which translates to $R_e \sim 2$ kpc for
$z=1-3$. This value is comparable to other studies of SMG in which the
sources were much better resolved, and so this gives a
measure of confidence in our fluxes and sizes \citep{Simpson2015,
  Hodge2016, Oteo2016, Rujopakarn2016,
  Fujimoto2017, Tadaki2017, Lang2019}. The limit for
reliable size determination in interferometric observations is given
by \citet{MartiVidal2012}, and using their Eqn.7 with $\beta=0.75$
and $\lambda_c=3.84$ the 2$\sigma$ limit to the minimum size which
could be reliably measured for our sources is listed in
Table~\ref{SMGT}. Two of the sources have fit results which are
comparable or slightly smaller than this limit and so for these we
quote the 2$\sigma$ upper limit to their sizes (SDP.5347,
SDP.6418). Figure~\ref{uvfitF} shows the azimuthally binned $(u,v)$
data, together with the model representing the best fit for the median
likelihood axial ratio, $a/b=0.75$.

There is a small difference between the peak flux derived from the fit
to the imaging and the 2D fit to the $(u,v)$ data described above
(where the size is a free parameter), such that the integrated fluxes
from the fits with finite size are higher than the point source
fluxes. This is expected, a point source flux estimate will always
underestimate the flux of a source with a finite size, the degree of
underestimation being dependent on the ratio of the source area compared to the
beam area. To be certain that this distinction (which for the most
part is not highly significant, the most significant being SDP.4104.s1
at 2.6$\sigma$) does not bias our subsequent study of the number
counts, we checked that repeating the analysis with the peak fluxes
instead of the integrated fluxes does not change the conclusions.

All fluxes and errors were then corrected for the primary beam
attenuation using the primary beam model output by CASA during the
{\sc clean} stage. The fluxes presented in Table~\ref{SMGT} are
corrected for the primary beam, and the correction made for the
primary beam attenuation is also listed there.

\section{An over-density of high redshift SMG around massive galaxies at $z=0.35$}
\label{smgS}

\begin{landscape}
\begin{figure}
\includegraphics[width=0.68\textwidth,trim=0.0cm 0.0cm 0.0cm 0.0cm, clip=true]{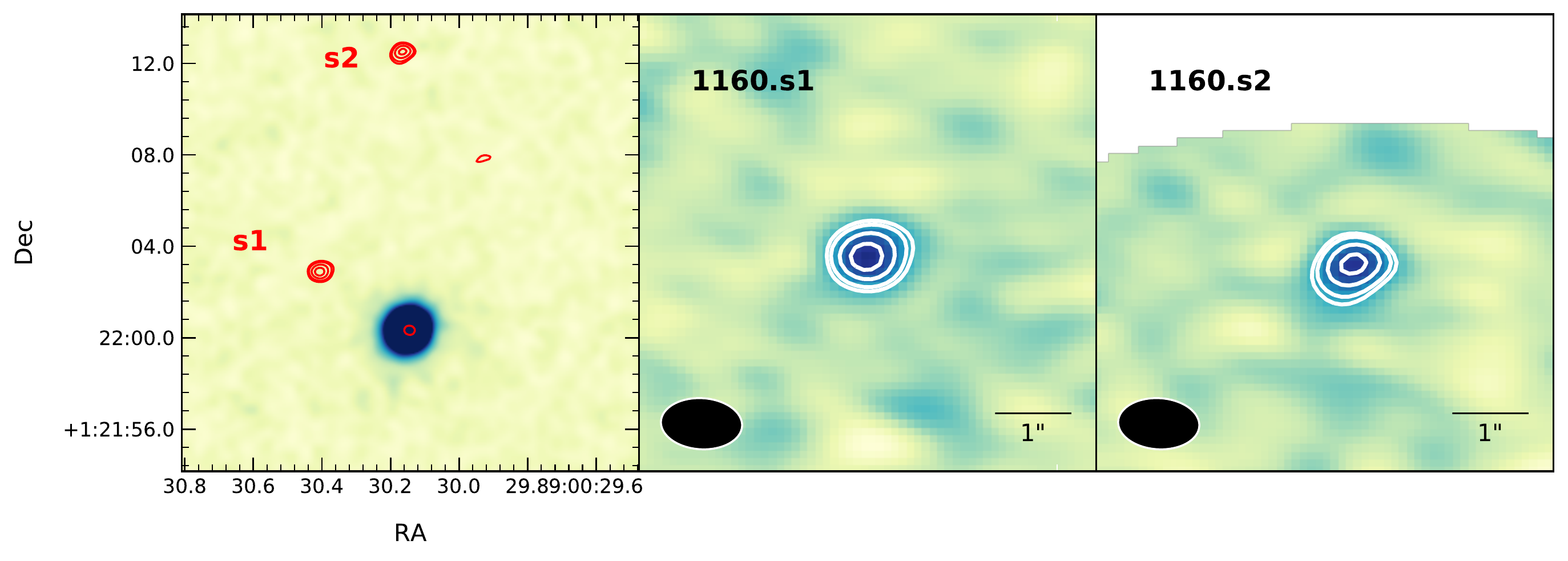}
\includegraphics[width=0.68\textwidth,trim=0.0cm 0.0cm 0.0cm 0.0cm, clip=true]{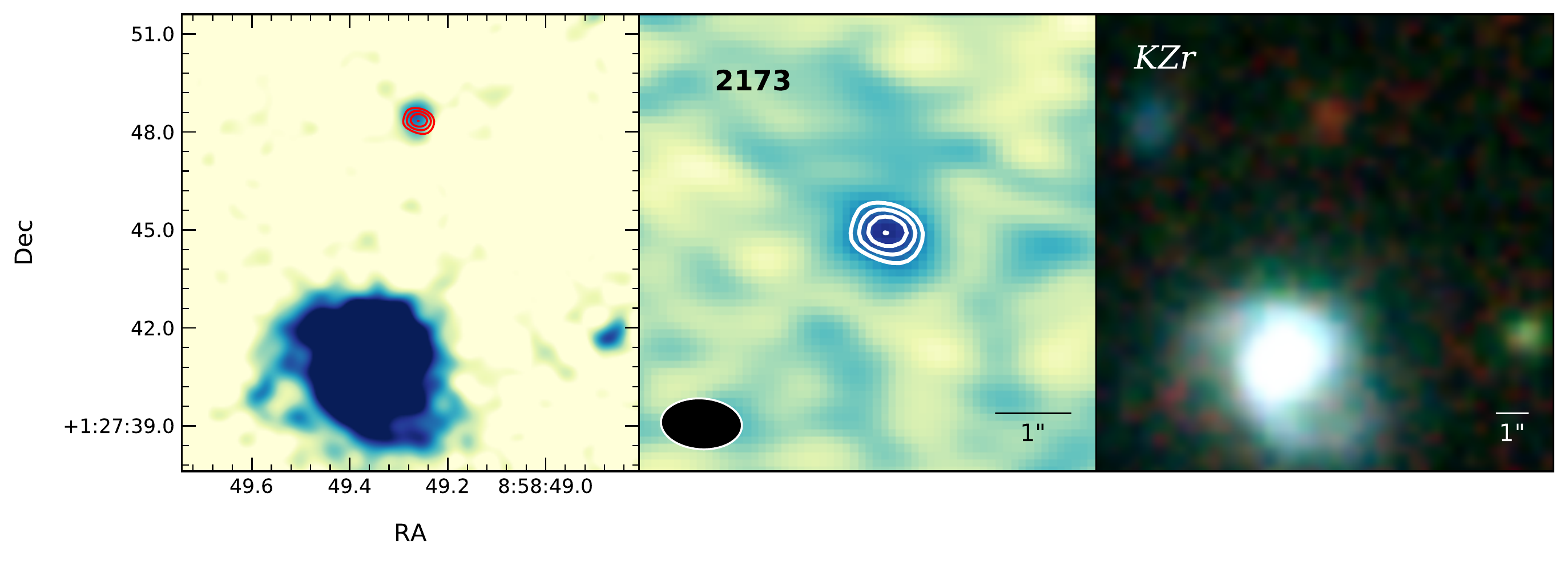}\\
\includegraphics[width=0.68\textwidth,trim=0.0cm 0.0cm 0.0cm 0.0cm, clip=true]{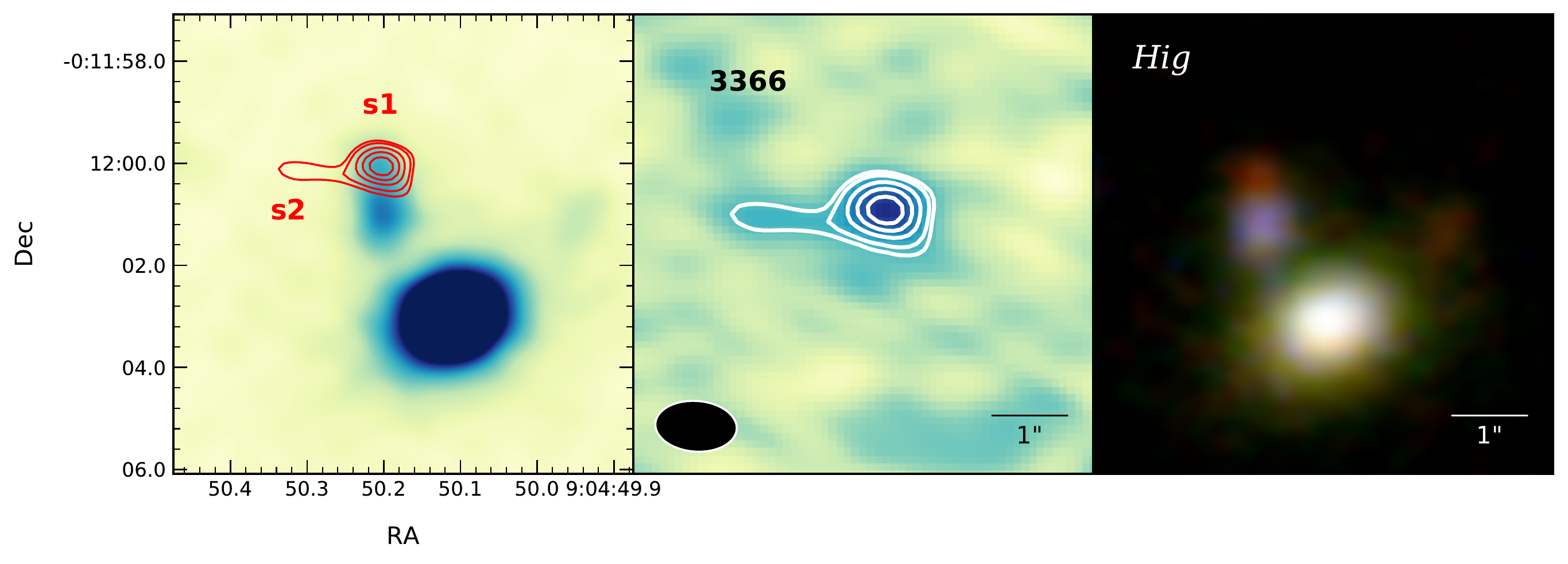}
\includegraphics[width=0.68\textwidth,trim=0.0cm 0.0cm 0.0cm 0.0cm, clip=true]{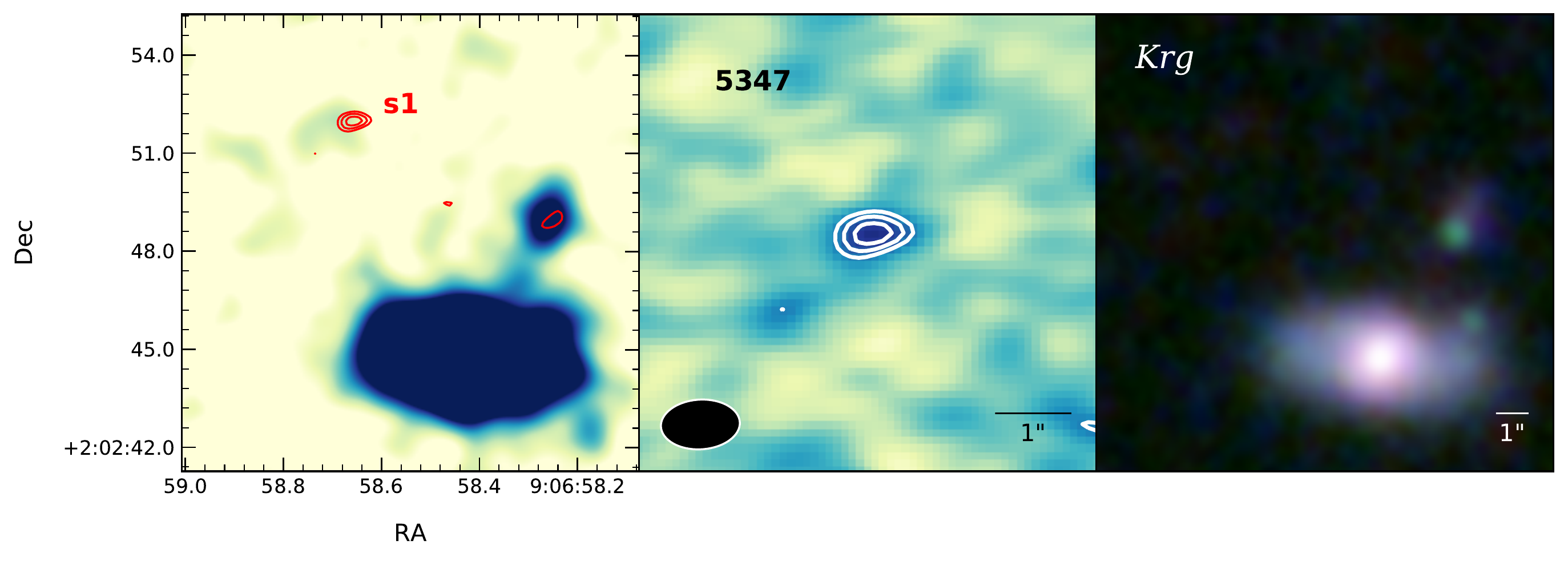}\\
\includegraphics[width=0.68\textwidth,trim=0.0cm 0.0cm 0.0cm 0.0cm, clip=true]{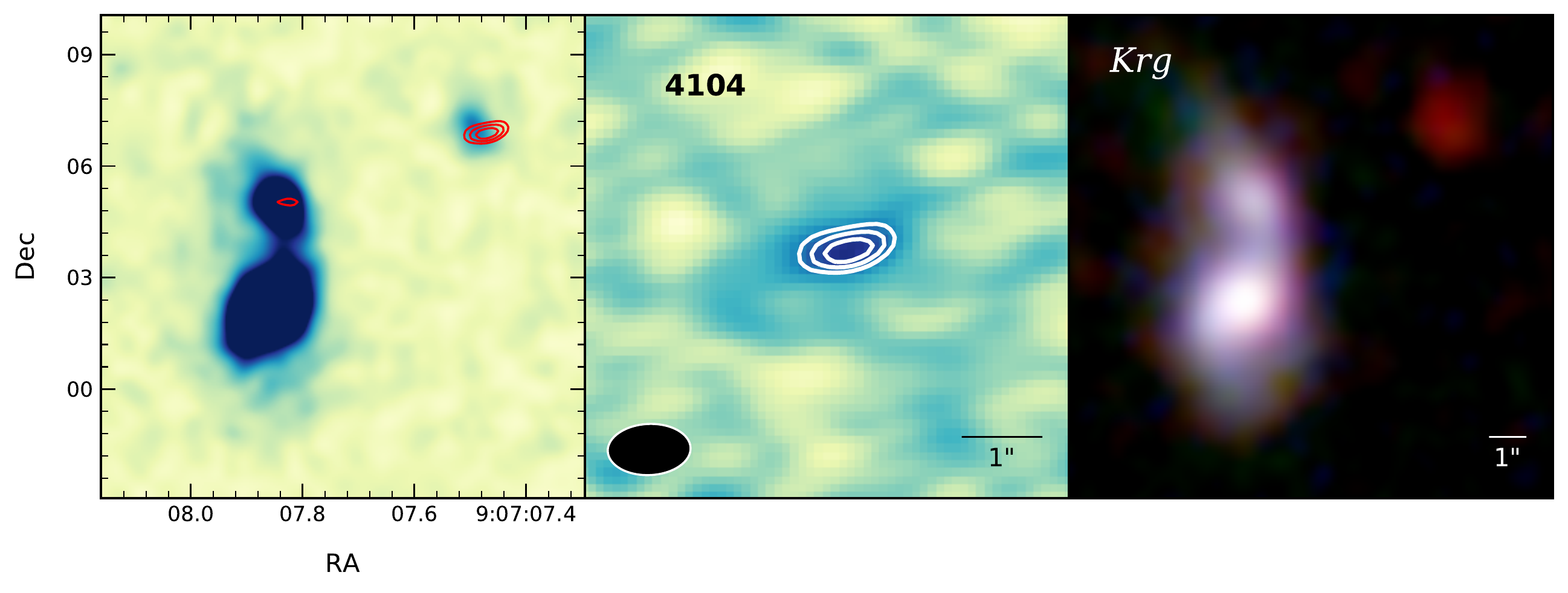}
\includegraphics[width=0.68\textwidth,trim=0.0cm 0.0cm 0.0cm 0.0cm, clip=true]{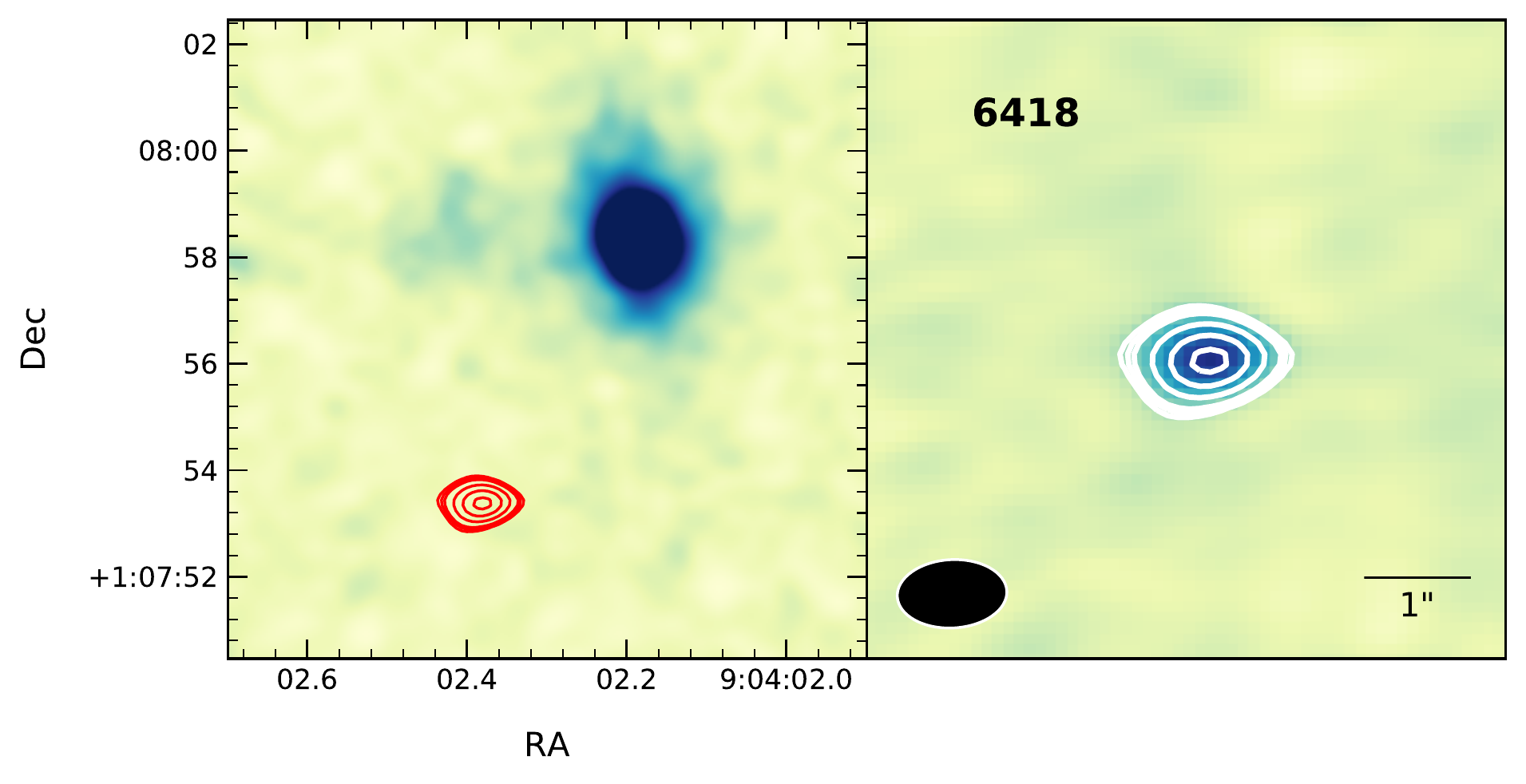}\\
\caption{\label{SMGF} Panels for each field containing SMG, showing
  from left to right: (Left) K-band image from VIKING \citep{Edge2013}
  with ALMA 850\mic contours overlaid in red, starting at
  4$\sigma$. The stretch is made to highlight any faint K-band
  emission from the position of the SMG. (Centre) ALMA 850\mic image
  zoomed in and centred on the SMG(s) with contours starting at
  4$\sigma$. Negative contours are shown starting at $-4\sigma$ as
  black dashed lines. (Right) For the four fields with K-band emission
  at the SMG position we show colour composites from VST KIlo Degree
  Survey \citep{deJong2017} and VIKING which highlight the very red
  colours of those sources. For SDP.1160 and SDP.6418, there is no
  sign of any K-band emission down to the VIKING 3$\sigma$ limit of
  $\rm{K_{AB}=21.3}$.}
\end{figure}
\end{landscape}

Half of the fields observed contained one or more serendipitously
detected high redshift SMG. In total 7 SMG are confidently detected at
$S_{850}>5\sigma$ in 6 fields. The details of these SMG are given in
Table~\ref{SMGT} and images are shown in
Figure~\ref{SMGF}.\footnote{The ALMA images have not been corrected
  for the variable attenuation of the primary beam. A source nearer to
  the edge of the map will have a higher flux at the same
  signal-to-noise contour level compared to a source in the centre. This is
  due to the fact that the primary beam attenuation increases the
  noise as a function of radius from the pointing centre.} We give
detailed notes on each source in Appendix~\ref{sourcesS}.

We looked for counterparts to these SMG in VIKING K-band imaging
\citep{Edge2013,Driver2016}, which is the deepest ancillary data-set in this
region, and find K-band counterparts to four out of seven
SMG. Counterparts or upper limits to the K-band magnitudes are noted in
Table~\ref{SMGT} and the K-band images are shown also in Figure~\ref{SMGF}.

\begin{table*}
\caption{\label{SMGT} Serendipitous robust SMG found in the fields surrounding $z=0.35$ H-ATLAS sources.}
\begin{tabular}{lccccccccc}
\hline
SMG  & R.A. & Dec. & $S_{850}$  & SNR & decon. size & $\theta_{\rm min}$ & $r$ & P.B. & $\rm{K_{AB}}$  \\
     &      &      &  (mJy)    &     & \asec & \asec  &      &     \\ 
\hline
1160.s1   & 09:00:30.40 & $+$01:22:02.90 & $1.06\pm 0.08$  & 11.4 & $0.4\pm0.09\times 0.3\pm0.08$ & (0.20)  & 5.0 & 0.82  & $>21.32$ \\
          &             &                  & $0.91\pm 0.08$  &     &    &   & &\\
1160.s2   & 09:00:30.17 & $+$01:22:12.44 & $4.05\pm 0.40$  & 11.1 & $0.34\pm 0.1\times 0.26\pm0.09$ & (0.21) & 12.5 & 0.20  & $>21.32$ \\
          &              &                 & $3.55 \pm 0.32$  &    &                             &        &   & \\
2173.s1   & 08:58:49.26 & $+$01:27:48.3  & $1.05\pm 0.11$   &  6.3 & $0.44\pm0.12\times 0.33\pm0.10$ & (0.28) & 7.5  & 0.59 & $21.45\pm0.03^{1}$ \\
          &             &                & $0.76 \pm 0.12$  &      &                              &      &        &   \\
3366.s1   & 09:04:50.20 & $-$00:12:00.10 & $0.82\pm 0.073$  &  11.0 & $0.39\pm0.10\times0.29\pm0.09$ & (0.21) & 3.4  & 0.91 & $19.02\pm0.03^{2}$  \\ 
          &             &                  & $0.74 \pm 0.067$ &       &       &         &  & \\
4104.s1   & 09:07:07.47 & $+$00:00:06.94 & $1.33 \pm 0.17$ & 6.0 & $0.6\pm0.19 \times 0.45\pm0.10$ & (0.29) & 7.2 & 0.58 & $20.65\pm0.12$ \\
          &             &                & $0.78 \pm 0.13$ &     &    &  &    &             \\
5347.s1   & 09:06:58.65 & $+$02:02:51.95 & $1.30\pm0.21$   & 7.4 & $0.25\pm0.22 \times 0.19\pm0.12$ & $<0.26$ & 8.0 & 0.52 & $21.55\pm0.31$  \\
          &             &                & $1.18 \pm 0.16$ &     &   &    &  &\\
6418.s1 & 09:04:02.38 & $+$01:07:53.4 & $2.56 \pm0.09$  & 23.7 &  $0.106\pm0.095 \times 0.08\pm0.065$  &  $<0.15$ & 6.0 & 0.75 & $>21.32$\\
        &             &               & $2.52 \pm0.07$  &     &                                     &   &       & \\ 
\hline
\end{tabular}
\flushleft{Name of SMG, R.A. and Dec., $S_{850}$: 850\mic flux from
  two methods. First row from fitting in the $(u,v)$ plane using {\sc
    uvmodelfit}, where the size fitted is also quoted. Second row is a
  minimum flux estimate, using the peak of a Gaussian fitted to the
  image with {\sc imfit}. SNR: signal-to-noise ratio from the peak
  pixel and rms map noise. $\theta_{\rm decon}$: Deconvolved size from
  the $(u,v)$ fitting (see text for details). $\theta_{\rm min}$ is the
  minimum reliable size which can be measured given the SNR of these
  sources using the formalism of \citet{MartiVidal2012}. Where the
  fitted size is comparable to or smaller than this size, we quote the
  2$\sigma$ size limit in this column. $r$: distance of the SMG to the
  centre of the target $z=0.35$ galaxy in arcseconds. P.B.: the primary
  beam correction which has been applied to these fluxes. $\rm{K_{AB}}$: K-band
  magnitude (AB) or 3$\sigma$ limit at the position of the SMG
  using VIKING data. $^1$ flux measured in a 2\arcsec aperture on
  VIKING image created by 
  \citet{Driver2016}. $^2$ flux measured on VIKING image using
  SeXtractor to deblend the red object from the foreground and using
  {\sc mag auto}.}
\end{table*}

\subsection{Over-density calculation}

In order to estimate 850\mic number counts from this sample, we must
first sum the area observed in which a given source could have been
detected with a SNR$>5\sigma$.\footnote{Given that our least
  significant source is actually 6$\sigma$, going down to 5$\sigma$
  provides a more generous area and therefore lower density estimate
  than if we had stated a cut off at 6$\sigma$.} 

The sensitivity of our ALMA pointings is not uniform, due to the
tapering effect of the primary beam. We created a noise-map for each of
the twelve fields to capture this radially varying noise
level by multiplying the primary beam attenuation image by the rms
noise measured in a source free region of the flux image before
primary beam correction. This gives us a map of the average noise as a
function of radius for each field. For the 12 noise-maps (one for each
of the 12 target fields), we sum the area over which we could have
detected each source at its peak measured flux at a significance level
of 5 times the local noise. The surface density of a source of a given
flux is then simply the inverse of this area. 

There are currently only two published measures of the faint number
counts at 850\mic using ALMA, those from \citet{Oteo2016} (henceforth
O16) who used the ALMA calibrator data-set to produce a direct and
unbiased measurement of the 850\mic counts with ALMA down to sub-mJy
levels, and more recently those from \citet{Bethermin2020} (henceforth
B20) who present serendipitous 850\mic sources detected in fields
targeted at $z>4$ [C{\sc ii}] emitters. Both of these analyses were
performed in comparable ways to ours, with a $\rm{SNR\geq5}$ selection
criteria and measurement of the source fluxes using integrated 2D
Gaussian fitting (although both of these works used image plane rather
than ($u,v$) fitting).

With such small numbers, we cannot hope to make an accurate measure of
the counts since our uncertainties are always going to be dominated by
counting statistics. Our aim in this paper is to determine with some
confidence level whether we are seeing a statistical over-density of
SMG in these fields containing massive $z=0.35$ galaxies. In order to
do this, we will compare our results to those of O16 and B20,
exploring the impact of different criteria for the flux measurement,
and the magnitude of completeness corrections. In each comparison, we
utilise the results of simulations by O16 and B20, who provide
completeness estimates for samples similarly selected with
$\rm{SNR\geq5}$.

To begin with, we provide short summary of the methods used by each of the
comparison counts analyses and highlight any points of difference.

O16 measure fluxes by fitting 2-D Gaussians to the image plane and
using the integrated flux, although sources are {\em detected} with
an initial peak pixel flux requirement $\rm{SNR\geq5}$. The beam-sizes
are $0.4\times0.3$\asec for 8 sources, and $1\times 0.6$\asec for 3
sources. Their simulations consist of point sources injected into the
visibilities and recovered using their normal procedure in a uniform
manner. These simulations demonstrate that completeness reaches
$\sim80$ percent at $\rm{SNR}=6$ for catalogues with a 5$\sigma$
detection threshold \citep{Oteo2016}. This is likely to be the best
case scenario as the simulated sources are all point-like while the
real sources would be expected to have finite sizes (particularly in the cases with
smaller synthesised beams).

B20 also measure their fluxes using integrated 2-D Gaussian fitting in
the image plane, having detected the source based on a peak pixel
$\rm{SNR\geq5}$. The beam sizes are typically $1.2\times 0.8$\asec. Despite
having a larger beam size compared to O16 (and other studies which
have measured the sizes of SMG), \citeauthor{Bethermin2020} infer that
the sources are marginally resolved by their $\sim 1$ \asec resolution
imaging, based on their finding that the ratio of the integrated flux
to the peak pixel flux is significantly greater than unity. While no
size measurements are explicitly mentioned in B20, we have used their
Eqn.4 and Fig. 4 which relate the peak/integrated flux ratio to
$\rm{\Omega_{source}/\Omega_{beam}}$, to estimate the average implied
{\em deconvolved} size of their sources as $\sim 0.72$\asec. The beam
size for our data-set is similar to B20, and yet our $(u,v)$ plane
analysis indicates that our sources have a size $<0.52\asec$, with a
mean of $0.31\asec$, comparable to the average size measured for SMG
in the literature ($\sim 0.3$\asec \citealp{Lang2019} and references
within). The upper range of the reported sizes in the literature is
$R_{e}\sim 4.5$ kpc \citep{Rujopakarn2016}, which translates to
FWHM$\sim0.68$\asec for redshifts greater than unity. It is
note-worthy that the {\em average} size of $\sim 0.72$\asec implied by
the B20 analysis of the peak-to-integrated flux ratio is larger than
the upper end of the range found in all of the rest of the SMG
literature. This has implications for our comparison of number counts,
because B20 perform their simulations of completeness including size
as a variable, finding that the larger sources have much lower
completeness at a given detection signal-noise ratio. In the B20
simulations, sources are injected directly into the image plane rather
than into the ($u,v$) visibilities as in O16. As a result, the
completeness corrections derived by B20 are very different at
the detection threshold $\rm{SNR=6}$ compared to those produced in O16 (80
percent for O16, 25--30 percent for a source with the average size of
0.72\asec implied by Fig. 4 in B20). Thus, for {\em a given set of
  sources, detected and measured in identical ways} the
\citeauthor{Bethermin2020} method would produce counts a factor $\sim
3$ larger than using the \citeauthor{Oteo2016} method, solely due to
the size dependence of the completeness correction, and the
interpretation of the peak/integrated flux ratio as a size measurement
by \citet{Bethermin2020}.\footnote{We note that B20 do not
  present any discussion of the impact of the source size on the
  derivation of the number counts, or the implications of their sample
  appearing to show such large extended sizes.}

Having contrasted the methods and data-sets used we now contrast the
number counts quoted by each survey. O16 use a fitting formula which
describes the counts all the way from 0.4~mJy to the bright-end
measured with ALMA by \citet{Simpson2015}.
\[
N_{850}(>S) = N_0\left[\left(\frac{S}{S_0}\right)^\alpha + \left(\frac{S}{S_0}\right)^\beta\right]^{-1}
\]
where $N_0=46.4\,\rm{deg}^{-2}$, $S_0=8.4$~mJy, $\alpha=1.9$ and
$\beta=10.5$. B20 do not fit a function to their counts but present
the cumulative counts in bins, which we logarithmically interpolate to
the flux values of interest to us.

The counts for the two analyses are presented in Table~\ref{countsT},
where we are listing the `robust $z<4$' counts from B20 to ensure we
are not including any over-density associated with the target sources.
The B20 counts are higher by a factor 1.7 -- 2.5 compared to those of
O16.\footnote{The errors quoted on the O16 counts are very much larger
  than 1/$\sqrt(N)$, but there is no explanation within the paper as
  to what other sources of error are contributing. Taking the O16
  errors at face value, the B20 counts are compatible within the
  1$\sigma$ uncertainty, but if we used a 1/$\sqrt(N)$ estimate for
  the O16 errors, then the B20 counts would be significantly higher.}
We must acknowledge two factors in the way the counts have been
obtained which may make the B20 counts tend to be higher than those of
O16. Firstly, the fields used to derive the B20 counts are targeted at
high-redshift galaxies with high star formation rates. They are
therefore biased sight-lines as galaxies are clustered, thus an extra
signal at the same redshift might be expected. For this reason, we
compare to the robust $z<4$ counts from B20 which will have removed
any contamination from clustered sources at the same redshift as the
targets. There is, however, another more subtle bias which may be
present, which is in fact the topic of this paper. Magnification
lensing bias means that the pre-selection of fields containing bright,
high redshift objects increases the probability of sight-lines
containing more large-scale structure; which weakly lenses the high
redshift sources making them appear slightly brighter, and hence more
likely to be selected as targets. Dusty galaxies present in any
foreground structures could create a bias to higher number
counts. Secondly, as mentioned earlier, the completeness corrections
adopted by each study are very different, with the size-dependent
completeness correction of B20 leading to much higher corrections on
average for their sources compared to the point source estimates of
O16.

Due to this uncertainty, we proceed to calculate our over-densities
with three different assumptions about the completeness correction:

\begin{enumerate}
\item{No completeness correction applied -- gives a minimum estimate of the counts for our fields.}
\item{Completeness correction following O16 simulations, meaning our two sources with $\rm{SNR\sim 6}$ (SDP.2173, SDP.4104) have corrections of 80 percent applied to them.}
\item{Completeness correction following B20 size dependent
  simulations, adopting the sizes measured for our sources using the
  $u,v$ fitting (see Table~\ref{SMGT}). This gives a completeness for SDP.2173 of 80 percent
  (as it is compact) but for SDP.4104, the largest source, completeness is only 35 percent}.
\end{enumerate}
 
In Table~\ref{countsT} we present the counts we measure with each of
the three completeness scenarios. We list the over-density of sources
in our survey relative to both the predicted number counts from
\citet{Oteo2016}, and relative to the robust ($z<4$) sample from
B20. The minimum over-density we find when we apply {\em no
  completeness correction} to our counts at $S_{850}>0.8, 1.3$~mJy, is
6.6--7.5 for O16 and 3--4 for B20. Furthermore, in contrast to the
other studies, we have not excluded the area in the centre of the map
which is covered by the target optical galaxy, {\em which makes our
  surface density estimates lower limits.}

Making a fairer comparison between our data and the literature, if we
compare our counts using the O16 completeness correction to the O16
counts, we find an over-density of a factor 7--8. If we compare our
counts using the B20 completeness correction to the B20 counts, we
find over-density factors of 5--5.5.  


We have estimated the 1-$\sigma$ uncertainties on our over-density
measurements using the Bayesian approach described by
\citet{Kraft1991}.  We used the same approach to estimate the
probability that we could find the observed number of sources in our
fields given the null hypothesis that there is no over-density
compared to the background counts.  Using the O16 counts, the
resulting probability rules out the null hypothesis at the 4-$\sigma$
level; using the B20 counts the null hypothesis is ruled out at
3.6-$\sigma$.  So, even though our over-density estimates have large
uncertainties, we are confident that there {\em is} a significant
over-density in our fields.

\begin{table*}
\caption{\label{countsT} The number counts estimated from our data compared to other ALMA counts at 850\mic.}
  \begin{tabular}{cccccccccc}
\hline
$S_{850}$  & $ n$  & Reference  & $N_{\rm obs}(>S)$& $N_{\rm cor}(>S)$& $N_{\rm bg}(>S)$& \multispan{2} Over-density  & P(null)$_{\rm obs}$ & P(null)$_{\rm cor}$ \\
 (mJy)   &        & counts     & (deg$^{-1}$) & (deg$^{-1}$) & (deg$^{-1}$)  &   observed &  corrected & ($\sigma$) &($\sigma$) \\
\hline
0.82 & 7 & O16 & 25735&  28089 &   3861 & $6.7^{+2.9}_{-2.2}$ &$7.3^{+3.1}_{-2.4}$ & 3.9 & 4.0 \\
1.3  &4  &           & 12109 & 13266  &  1609 & $7.5^{+4.5}_{-3.2}$ &$8.2^{+4.9}_{-3.5}$ & 3.1 & 3.2 \\
\hline

0.82 &7 & B20 & 25735 & 35529 &   6431 & $4.0^{+1.7}_{-1.3}$ &  $5.5^{+2.4}_{-1.9}$  & 3.1 & 3.6 \\
1.3  &4 &                & 12109 & 20707 &   4033 & $3.0^{+1.8}_{-1.3}$ &  $5.1^{+3.1}_{-2.2}$ & 2.1 & 2.7\\
    \hline
  \end{tabular}
\flushleft{\small Estimated counts in bins of 850\mic flux. $n$ is the
  number of sources per bin, $N_{\rm obs}(>S)$ is the surface density we
  measure {\em not corrected} for completeness, $N_{\rm cor}(>S)$
  accounts for the completeness correction using the same method as
  the background counts being compared to. $N_{\rm bg}(>S)$ are the
  background cumulative counts from \citet{Oteo2016} (O16) and
  \citet{Bethermin2020} (B20) respectively. These {\em are corrected}
  for completeness. The over-density columns are the relative
  over-density in our fields (both uncorrected and corrected for
  completeness) compared to the literature `blank-field' measures
  including 1$\sigma$ errors derived from the approach of
  \citet{Kraft1991}. P(null) is the significance at which the null
  hypothesis is rejected (that the number of sources we see is a
  random realisation of the number counts given by the literature
  reference). }
\end{table*}

We next consider a possible physical explanation for this excess of SMG.

\subsection{Lensing Magnification Bias}
\label{lensS}

Cross-correlations between foreground structure traced by SDSS
galaxies and background galaxies detected by {\em Herschel} have been
detected with high significance \citep{GN14,GN17}. Anomalies in the
positional offset distribution between {\em Herschel}-ATLAS sources
with high-redshift submillimeter colours and foreground optical
galaxies are also seen \citep{Bourne2014}. These findings both imply
that there is significant magnification of the background SMG
population by large scale structure in the foreground. The recent work
by \citet{GN17}(hereafter GN17) shows that lensing from the foreground
halo will produce an over-density of SMG relative to the unlensed
background within $r< 20$\asec of halos with $\rm{log\,M_h/\msun >13.5}$. The
magnitude of the over-density will correlate with the distance of the
SMG from the projected centre of the halo mass distribution.

\begin{figure}
\includegraphics[width=0.5\textwidth,trim=0.5cm 0.5cm 0.0cm 0.0cm,
  clip=true]{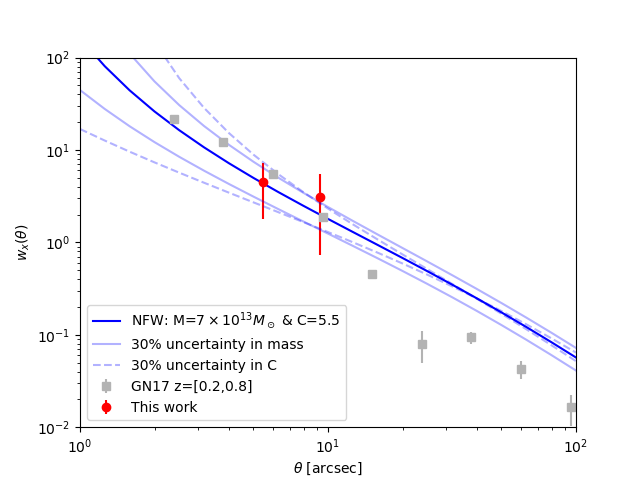}
\caption{\label{xcF} The cross correlation signal for background SMG
  and foreground structure in the range $0.2<z<0.8$ from SDSS and
  GAMA. Grey points are the results of the statistical
  cross-correlation using H-ATLAS sources with estimated FIR photo-z
  $z>1$ from GN17. Red points are the SMG found in the fields of our
  $z=0.35$ targets in two radial bins. The solid blue line is for a
  NFW halo with $M_{200c}=7\times10^{13}\,\msun$ and concentration
  parameter $C=5.5$. The dashed lines represent the model with a
  variation of $\pm30$ percent in $C$, while the translucent solid
  lines show a variation of $\pm30$ percent in halo mass.}
\end{figure}

We have computed the relative over-density of SMG as a function of
radius from the centre (assuming that the target $z=0.35$ galaxy is
the centre of its halo). Figure~\ref{xcF} shows our estimated
cross-correlation (red dots) compared to the cross-correlation
measurements obtained by GN17 (grey squares), who studied the
magnification bias due to Luminous Red Galaxies with $0.2<z<0.8$ from
GAMA \citep{Baldry2018} which act as lenses on the high redshift
($z>1.2$) SMG detected by H-ATLAS. There is a good agreement between
both set of measurements. To derive physical information on the
typical halo that can produce such magnification bias, we perform a
similar analysis to that described in \citet{Bonavera2019}. By
neglecting the shear effect in the magnification bias,
\citeauthor{Bonavera2019} exploit the direct relationship between the
cross-correlation function and the halo convergence:
$\kappa(\theta)=1-(w_x(\theta)+1)^\frac{-1}{2(\beta-1)}$, where
$\beta\gtrsim 3$ is the slope of the integrated source counts of the
background SMG sample. As is common in previous works analysing the
magnification bias, we model a Navarro Frenk and White (NFW: \citealp{nfw})
mass density profile with the mass, $M_{200c}$, and concentration,
$C$. The data do not allow a direct constraint with the current
statistics, but Figure~\ref{xcF} shows that they are consistent with
typical values of $M_{200c}=7\times 10^{13}\, \msun$ and $C=5.5$ (blue
solid line). Varying each parameter by $\pm30$ percent produces the
dashed lines ($C$) and the pale blue solid lines ($M_{200c}$). There
is intrinsic degeneracy between mass and concentration in the fitting,
and to provide direct constraints will require a larger sample which
could be split over more radial bins.

The values which are representative of the cross-correlation are in
good agreement with the $M_{200c}$ vs $C$ relationships from
\citet{Child2018} and \citet{DuttonMaccio2014}. The magnifications
expected at $r=3-12\asec$ for halos this massive are of the order
$\mu=1.5-3$ \citep{GN14} which is enough to make a substantial
over-density of SMG visible due to the steep number counts in the
sub-mm waveband. This halo mass corresponds to $M_\star\sim 1-3\times
10^{11}\, \msun$ using the relationships for $M_{200c}-\Ms$ from both
\citet{Moster2010} and \citet{Behroozi2013}. We might expect to find a
small group of galaxies consisting of a bright, massive central galaxy
with a few additional dwarf satellites of much lower mass.

The six SMG fields have central galaxies with $\Ms =
6.5\times10^{10}-1.6\times10^{11}\,M_\odot$; four of the six central
galaxies are interacting with smaller satellites (evidence for this is
the optical morphology and kinematics of CO and \CI, see Dunne et
al. {\em submitted} for details), and a fifth (SDP.1160) is a member of
a GAMA spectroscopic group with another $r<19.8$ galaxy lying outside
the ALMA B7 field of view. The only galaxy for which we have no
spectroscopic evidence for a close neighbour is SDP.2173, although
there are three dwarf galaxies in the KIDS catalogue
\citep{deJong2017,KIDSMLPz,Wright2018} which have $z_{ph}=0.37-0.38$
located 40-90 kpc in projection and could very plausibly be in the
same halo. SDP.2173 also has the brightest r-band magnitude of any of
the sources we observed, $R_{AB}=-21.70$. {\em All} four of the
interacting systems in our sample of 12 have an SMG in the field. The
GAMA survey is highly spectroscopically complete for $r<19.8$
galaxies, and groups with $N\geq 2$ spectroscopic members have been
catalogued in the GC$^3$ group catalogue of
\citet{Robotham2011}. While only SDP.163 and SDP.1160 are listed as
group members in the GC$^3$, our own exploration of the GAMA data cube
reveals a galaxy (GAMA 372510) located a co-moving projected distance
of $D=0.86$~Mpc with a $dv=396\kms$ from SDP.6418. Not all the
$z=0.35$ galaxies with SMG in the field would be expected to be
designated as groups by GAMA for two reasons. Firstly, the GAMA
criteria for group identification requires $N\geq 2$ sources with
$r<19.8$ meeting the radius and velocity criteria. Smaller groups with
one dominant bright galaxy would therefore probably not be identified
as such because the second brightest group member would be fainter
than the GAMA spectroscopic limit. All of our ALMA targets have
$r<19.4$, with a median $r=19.0$ for the fields containing SMG -- much
brighter than the typical GAMA group brightest galaxy at this
redshift. Secondly inspection of the GAMA N(z) shows that the
sensitivity to large scale structure drops dramatically at $z>0.3$ so
the group catalogue is likely to be incomplete for intermediate mass
groups at the redshifts of this sample.  

The environments and optical properties of the SMG field target
galaxies are listed in Table~\ref{environT}.

\begin{table}
\caption{\label{environT} Environment of the $z=0.35$ galaxies which have SMG in the field.}
\begin{tabular}{lccc}
\hline Field & $\rm{R_{tot}}$ & Environment & dwarf $z_{ph}$\\ \hline
1160 & $-21.98$ & GAMA group (103864) & \\ 2173 & $-21.70$ & 3 dwarfs
$r_p=40-90$ kpc & 0.37--0.38\\ 3366 & $-21.65$ & close pair &
0.26--0.38\\ 4104 & $-21.37$ & close pair & 0.37\\ 5347 & $-21.39$ & 2
dwarfs, signs of interaction & 0.62\\ 6418 & $-22.04$ & GAMA 372510:
$r=0.86$~Mpc, & 0.42\\ & & and $\Delta v = 396$\kms & \\ & & also
dwarf interaction & \\ \hline
\end{tabular}
\flushleft{$\rm{R_{tot}}$ Total absolute r-band magnitude of the
  galaxies in the halo. Environment: commentary on the type of
  environment for the target $z=0.35$ galaxy whose halo is responsible
  for the lensing magnification bias. Dwarf $z_{ph}$: photo-z of any
  dwarf satellite galaxies using KIDS data
  \citep{deJong2017,KIDSMLPz,Wright2018}.}
\end{table}

\section{Boosting of the H-ATLAS fluxes by SMG in the beam.}
\label{boostS}

Lensing by the halo of the massive $z=0.35$ system
increases the probability of finding a high redshift SMG within the
same {\em Herschel} beam as the target low redshift source. This is a
proverbial can of worms for modelling statistical properties of sub-mm
surveys, e.g. source counts, sub-mm colour distributions and SED
modelling. Flux boosting from the SMG will affect all SPIRE fluxes, but
those at longer wavelengths disproportionately so for two reasons.
Firstly, the {\em Herschel} beam is larger at longer wavelengths, thus
a contaminant at distance $d$ will have a higher beam profile
weighting at longer wavelengths. Secondly, the sub-mm colour
of the high redshift SMG will be redder (relatively brighter at
500\mic) compared to the target source, as the observed frame samples
closer to the peak of the SED at high redshift. The percentage contamination to the 500\mic flux
from the high-z SMG is thus larger than that at 250\mic.

To estimate the contamination of the H-ATLAS fluxes for this sample at
$z=0.35$, we take the ALMA 850\mic flux measured for each SMG ($S_{\rm
  SMG}$) and calculate the SPIRE fluxes the SMG should have using
plausible values for the dust SED. We consider a range of redshift
($1<z<5$), ruling out redshift ranges where the predicted 250--500\mic
signal would be highly visible in the H-ATLAS maps at the position of
the SMG. For the remaining possible redshifts we calculate
$S_{\lambda}^C$, the contribution of the SMG to the {\em Herschel}
$\lambda$\mic fluxes of the $z=0.35$ galaxy, by weighting the predicted SMG flux in
each SPIRE band by the beam attenuation at the location of the
$z=0.35$ galaxy relative to the position of the SMG:
\[ 
S_{\lambda}^{\rm C} = S_{\rm SMG} \left(\frac{850\mu m}{\lambda}\right)^{4+\beta}\frac{e^{h(1+z)/c\lambda_{850} kT_d}-1}{e^{h(1+z)/c\lambda kT_d}-1}\,e^{-d^2/0.36 \theta_{\lambda}^2}
\] 
Here $\lambda$ refers to the SPIRE wavelength of interest in \mic,
${\theta_{\lambda}}$ is the relevant beam size for SPIRE (18\asec,
25\asec and 35\asec for the 250, 350 and 500\mic bands). $S_{\rm SMG}$
is the 850\mic ALMA flux of the SMG and $z$ is the redshift chosen for the
prediction. The dust SED parameters we use are $T_d=38$~K and
$\beta=1.8$, which are plausible values for SMG
\citep{Chapman2005,daCunha2015,Stach2019}, and $d$ is the separation
between the SMG and the centre of the target galaxy in arcsec.

To determine the most likely contamination values, we step through in redshift from
1.5--5 in steps of $\Delta z=0.5$, calculating the estimated
contamination signal at each redshift. We adjust the H-ATLAS catalogue
fluxes for the low redshift targets (Table~\ref{photomT}) by these contamination values and
then fit the SED from UV--850\mic using {\sc magphys} (see
Section~\ref{sedsS}). We choose the redshift ($\rm{z_{SED}}$) at which
the correction produces the lowest overall $\chi^2$ for the SED fit;
$z_{\rm SED}$ and contamination values are listed in
Table~\ref{SMGcontamT}.\footnote{The redshift and \td are roughly
  degenerate in this process but we are not attempting to constrain
  either; merely we wish to retrieve FIR colours which are most
  compatible with the PACS and ALMA 850\mic photometry, which we know
  to be uncontaminated.} Interestingly, we find that the fields
containing SMG with K-band counterparts produce the best
fits when the SMG is assumed to be at $z=1.5-2$ while those without
K-band counterparts have better fits when the SMG is assumed to have
$z>3$.

\begin{table}
\caption{\label{SMGcontamT} Estimated sub-mm fluxes for the SMG and
  their boost effect to the $z=0.35$ H-ATLAS source photometry.}
\begin{tabular}{lcccc}
\hline
SMG & $\rm{z_{SED}}$ & $\rm{S_{250}}$ & $\rm{S_{350}}$ & $\rm{S_{500}}$ \\
   &                &  (mJy)        &   (mJy)       & (mJy) \\
\hline
1160.s1  & 2 & 11.0 (9.0) & 8.0 (7.1) & 3.9 (3.7)\\
1160.s2  & 3  & 16.4 (4.6) & 18.1 (9.0) & 12.0 (8.6)\\
1160.H1  & 4  & 17.0 (0.02) &  28.6 (0.8) & 26.0 (4.5) \\
2173     & 1.5 & 14.4 (9.2) & 8.6 (6.7) & 3.7 (3.3)\\
3366     & 1.5 & 13.7 (12.5) & 8.1 (7.7) & 3.5 (3.4)\\
4104     & 2 & 19.3 (12.7) & 14.0 (11.1) & 6.9 (6.2)\\
5347     & 4 & 3.3 (1.2) & .. (2.5) & ... (2.7)\\
6418     & 5 & 1.6 (1.2) & 4.1 (3.5) & 5.1 (4.7) \\  
6451.b7  & 4 & 1.2 (0.7) & 1.9 (1.5) & 1.8 (1.5)\\
6451.b3  & 4 & 3.4 (0.9) & 5.7 (2.7) & 5.2 (3.6) \\
       
\hline
\end{tabular}
\flushleft{\small $\rm{z_{SED}}$ is the redshift used to compute the SMG
  fluxes for an SED with $T_d=38$~K, $\beta=1.8$ and the measured
  $S_{850}$ flux from Table~\ref{SMGT}. Predicted flux of the SMG in the
  {\em Herschel} bands and in parentheses its contamination to the H-ATLAS
  fluxes of the $z=0.35$ galaxy. The 1160.H1 red source catalogue fluxes, 18.0, 27.7, 23.9 mJy at
  {250, 350, 500\mic} \citet{Ivison2016} are replicated with an
  $S_{850}=10$~mJy SMG for our standard SED, which produces the
  contamination listed above. 6451.b7 is a 4.9$\sigma$ source
  in the band 7 image with a flux of $S_{850}=0.68$ mJy. 6451.b3 is a
  potential line emitter at 3~mm, which is given the maximum
  $S_{850}<2$~mJy based on the 3$\sigma$ noise at its position in the
  B7 image.}
\end{table}

The boosting effect will also mean that a low redshift sub-mm source
is more likely to be above the detection threshold in a flux-limited
survey if its halo is acting as a lens for a SMG with small projected
separation. In order to see what effect the best estimate of our boosting has on our
initial sample selection, we subtract the 250\mic contamination in
Table~\ref{SMGcontamT} from the flux reported in the H-ATLAS SDP
catalogue from \citet{Rigby2011}.\footnote{This is the relevant
  catalogue for the purpose of this calculation since it was from here
  that the sample was originally selected.} We then determine how many
of the sources would still have been in the catalogue at
$S_{250}/\sigma_{250}\geq5$ had there been no SMG. There is
clearly a lot of uncertainty in this rough estimate because a wide
range of contaminating 250\mic fluxes still produce acceptable fits to
both the {\em Herschel} maps and the SED fits. The two brightest
sources with SMG in the field (SDP.1160 and SDP.2173) remain at
$\rm{S_{250}^{cor}>5\sigma_{SDP}}$ for any reasonable SMG
contamination estimate. Of the other four, there are possible redshift
ranges where the contamination would still allow them to remain
in the sample (e.g. $z>2.5$ for SDP.3366 and SDP.4104, $z>3$ for
SDP.5347 and $z>5$ for SDP.6418). 

Depending on the redshift of the SMG, it is entirely possibly that all
six source would remain in the SDP sample even without the presence of
the SMG, for the best estimate of contamination in
Table~\ref{SMGcontamT} 4/6 would remain in the sample and in the worst
case, pushing the contamination to the highest permitted values only
2/6 would remain. 

The possibility that the high fraction of `lensing' systems in our
250\mic catalogue is higher than it should be due to the effect of the
boosting does not negate the magnification lensing bias as an
explanation for this effect, however it would need to be accounted for
in the modelling to derive more accurate parameters. This is beyond
the scope of this study, and much larger samples which could be selected in
flux bins to mitigate this uncertainty would be required to exploit this further.

\subsection{How much boosting is there for flux limited surveys with {\em Herschel}?}

The sample of 12 galaxies at $z=0.35$ targeted by ALMA is a blind sample of 250\mic
selected sources from H-ATLAS, and while small, it is an unbiased set
of sources from that survey. The implications of so much boosting in
the long wavelength H-ATLAS photometry could be profound. Taking the
results from Tables~\ref{photomT}\&~\ref{SMGcontamT} we find that the
average 350 (500)\mic boost for the fields with SMG is 1.44 (1.75). If
we assume the other fields have no contamination at all then we arrive
at a global average boost factor of 1.26 (1.44) at 350 (500)\mic. This
is significantly higher than that estimated from the simulations for
the H-ATLAS data release (DR1) by \citet{Valiante2016}, who report an
average boost of 1.09-1.1 for these flux densities in the 350 and
500\mic bands. At 250\mic the average boost is 1.23 for the SMG fields
and 1.13 overall (assuming the other fields have a boost of 1.0). This
agrees reasonably well with the \citet{Valiante2016} and
\citet{Rigby2011} simulations indicating that the analysis of 250\mic
data from H-ATLAS is tractable using the corrections produced in the
data release papers. This boosting by lensing bias will be a redshift
dependent effect because there is a higher probability for lensing
events to occur for halos within a certain redshift range (dependent
on the redshift distribution of the background population). For
typical SMG, simulations by \cite{lapi12} suggest that this lensing
bias will be greatest for lens halos around $z\sim 0.5$ and will
decrease sharply below $z=0.1-0.2$. To determine the boosting
corrections with more accuracy, a larger sample across a larger
redshift range would be required. This effect will not be limited to
the H-ATLAS survey but will affect any {\em Herschel} survey where the
sources are in the redshift interval with a high probability for
lensing ($0.2<z<0.7$). The method of flux measurement is also unlikely
to mitigate these boosts unless any high redshift SMG are identified
at other wavelengths during source extraction using a
cross-identification (XID) method
\citep[e.g.][]{hurley17,pearson17,wang19}

\subsection{Impact of ALMA data on SED fits to {\em Herschel} photometry}
\label{sedsS}
The discovery of such a large boost to the 350\mic and 500\mic fluxes
in this sample suggests that in the absence of ALMA information, SED
fits to H-ATLAS (and any other non-deblended {\em Herschel})
photometry could be biased in some significant fraction of sources.

To assess this, we fitted UV-FIR SEDs for the $z=0.35$ targets to two
versions of our photometry. We used the energy balance SED fitting
code {\sc magphys} \citep{DaCunha2008} which uses libraries of optical
and infrared SEDs with parameters drawn stochastically from physically
motivated priors. We use extended infrared libraries which have an ISM
cold temperature range of 10--30~K, as several of the sources favoured
warmer ISM temperatures than allowed in the standard {\sc magphys}
infrared libraries. More details of the ancillary data used in the
fitting and the full results for the $z=0.35$ sample are in Dunne et al. {\em submitted}.

In Case 1, we used H-ATLAS only data for the FIR photometry
(Table~\ref{photomT}: longest wavelength 500\mic) without any
adjustment for the contamination by high-z sources, i.e. what we would
have done in ignorance of the 850\mic information from ALMA. In Case 2,
we used our full FIR data-set including the ALMA 850\mic flux and
having corrected the SPIRE 250--500\mic photometry for the
contamination by any high-z SMG in the field as described in
Section~\ref{boostS}.

The results of the comparison are shown in Table~\ref{SEDcompT}, where
we compute $\lsub^{\rm SED}$ from the {\sc magphys} fit as has been
done in some literature studies \citep{Hughes2017}.  The fits using
only the {\em Herschel} photometry give $\lsub^{\rm SED}$(500\mic)
while those which additionally include the ALMA B7 data give
$\lsub^{\rm SED}$(850\mic).

The average offset in the SED-based 850\mic luminosity, $\Delta
\lsub^{\rm SED}=\lsub^{\rm SED}(500\mu m)-\lsub^{\rm SED}(850\mu m) =
0.15$ dex (for the fields without SMG the average is 0.09 dex, while
for those fields with SMG the average is 0.2 dex). The temperatures
estimated for the cold dust in the diffuse ISM are sensitive to the
{\em Herschel} SPIRE colours, and as expected, these are also biased in
the fits to H-ATLAS only photometry in the fields with SMG. In
Table~\ref{SEDcompT}, $T_{500}$ is the cold ISM temperature from {\sc
  magphys} for the fits to {\em Herschel} only photometry, while
$T_{850}$ are the temperatures from fits including the ALMA
photometry, and having corrected the Herschel fluxes for SMG
contamination. The average temperature without ALMA photometry is
$\langle T_{500} \rangle = 22.2$~K (median 21.6~K), while that with
ALMA 850\mic data and corrected fluxes is $\langle T_{850} \rangle =
24.4\pm0.8$~K (median 24.5~K). Part of this effect is
  independent of any boosting by SMG and is due to the lower
  signal-to-noise at the 350 and 500\mic wavelengths compared to
  250\mic in the {\em Herschel} photometry. The long-wavelength ALMA
flux constrains the cold dust temperature and mass, and without it we
suspect that {\sc magphys} fits are returning close to the median of
the flat prior (20~K).


For the dust masses the bias is greater with an average offset
$\Delta\Md=\Md(500\mu m)-\Md(850\mu m) = 0.24$, again the effect is
much larger in fields with SMG (0.36 vs. 0.11). The effect on the
derived dust mass from {\sc magphys} is larger than that on
$\lsub^{\rm SED}$ because the dust mass is sensitive to temperature as
well as the overall flux, and the temperatures from the {\em
  Herschel} only fits are lower than those which include the ALMA data.

The effect of the flux boosting on the SED fits is much smaller {\em
  once ALMA data are present} because the ALMA 850\mic data force such
a tight constraint on the SED shape. Using uncorrected SPIRE fluxes in
conjunction with ALMA 850\mic data produces differences in SED
parameters which are well within the 1$\sigma$ errors ($\Delta < 0.03$
dex in \Lir, $\Delta < 0.1$ dex in \Md, $\Delta < 0.1$ dex in sSFR,
and $\Delta < 1$~K in $T_c$). Therefore, the boosting corrections are
not so important as long as the high resolution long wavelength data
are present; however without ALMA data, the lack of boosting
correction may lead to large biases in analyses of {\em Herschel}
samples. For example, the VALES survey \citep{Villanueva2017} was an
ALMA CO(1-0) follow-up of 160\mic selected H-ATLAS sources in the
redshift range $0.1<z<0.4$. In one paper \citep{Hughes2017} the
H-ATLAS photometry was fitted using {\sc magphys} and the SED
extrapolated to give $\lsub^{\rm SED}$(500\mic). This was then
compared to the CO luminosities and calibrations for the dust-to-ISM
mass factor \asub derived. However, in a re-analysis of the literature,
Dunne et al. {\em in prep} show that the average \lsub/\lcoa from
VALES is 0.2 dex lower than {\em any} other sample of low or high
redshift sources, including this $z=0.35$ sample. We expect that this
offset in \lsub/\lcoa is due to the lensing magnification boosting
effect shown in this work; Table~\ref{SEDcompT} shows that the
difference in $\lsub^{\rm SED}$ inferred with and without the ALMA
data is of this order. Other analyses potentially affected by this
bias are the evolution of the 350 and 500\mic luminosity and dust mass
functions \citep{Dunne2011,marchetti16}, modelling of the 350
and 500\mic source counts
\citep{clements10,Bethermin2012,Valiante2016}, SED evolution
\citep[e.g.][]{symeonidis13,viero13} and stacking analyses
\citep[e.g.][]{Bourne2012,viero13,schreiber15}. In particular, the
stacking analyses of both \citet{Bourne2012} and \citet{viero13} found
that the dust temperatures decreased with stellar mass in the lowest
redshift bin (as we would predict based on the extra boosting at 350
and 500\mic due to the cosmic lensing bias). In \citet{Bourne2012}, a
strong increase in the stacked 500\mic luminosity of optically red
(passive) galaxies was also seen from $z=0-0.35$. A strong lensing
explanation did not seem to explain those findings at that time, but
the possibility of the lensing bias reported here and in \citet{GN14}
was not known at that time. A strongly evolving optically red population which
had optical spectral signatures (in stacks) of both old stellar
populations and ongoing SF was found in \citet{Eales2018} using the
H-ATLAS catalogues. Could these optically red sources with strongly
evolving dust emission be sign-posting the halos producing the
magnification lensing bias?

A larger sample of $0.2<z<0.7$ galaxies selected at 250\mic and imaged with ALMA will be required to
understand and address these issues, and to determine the statistics
of boosting due to this effect in bins of redshift and stellar mass.

\begin{table}
\caption{\label{SEDcompT}} The difference in SED fitted parameters when using {\em Herschel} only photometry compared to the corrected {\em Herschel}$+$ ALMA data.
\begin{tabular}{lcccccc}
\hline 
Name & SMG & $T_{850}$ & $T_{500}$ & $\Delta \Md$ & $\Delta \lsub^{\rm SED}$ \\ 
\hline
163 & N & 22.5 & 21.6 & 0.07 & 0.08 \\
3132 & N & 25.9 & 22.4& 0.19 & 0.15 \\
5323$^{\dag}$ & N & 21.8 & 16.5 & 0.53 & 0.45\\ 
5526 & N & 24.4 & 25.6 & $-0.10$ & $-0.08$ \\
6216 & N & 20.0 & 20.3 & $-0.05$ & $-0.1$ \\
6451 & N & 25.0 & 25.7 & 0.03 & 0.06 \\
\hline 
1160 & Y & 28.3 & 21.5 & 0.66 & 0.32\\ 
2173 & Y & 24.5 & 19.7 & 0.40 & 0.26\\ 
3366 & Y & 26.5 & 20.2 & 0.71 & 0.58\\ 
4104 & Y & 27.7 & 25.4 & 0.22 & 0.06\\ 
5347 & Y & 19.9 & 20.4 & 0.03 & 0.02\\ 
6418 & Y & 23.4 & 21.7 & 0.14 & 0.001\\ 
\hline 
\end{tabular}
\flushleft{\small{Notes: Sources are grouped as to whether they have
    an SMG detected in the field at 850\mic. $T_{850}$ is the cold
    dust temperature fitted by {\sc magphys} using the 850\mic ALMA
    data and making corrections for blending as described in
    Section~\ref{boostS}. $T_{500}$ is the temperature fitted to the
    catalogued H-ATLAS photometry only with no corrections. $\Delta
    \Md$ is the difference in log \Md between the H-ATLAS only and
    corrected H-ATLAS+850\mic photometry. $\Delta \lsub^{\rm SED}$ is the difference in
    log $\lsub^{\rm SED}$ from the best fitting {\sc magphys} SED in the H-ATLAS
    only and corrected H-ATLAS+850\mic cases. $^\dag$ SDP.5323 has no PACS
    coverage and therefore its SED is very poorly constrained without
    the ALMA data point. We do not use this source in the analysis of
    average properties.}}
\end{table}

\subsection{Summary of flux boosting and its implications}

In summary, the 250\mic flux is the least affected of the SPIRE bands
by the presence of a background SMG because (a) the beam size
increases with increasing wavelength, and (b) the K-correction favours
means that high redshift sources are relatively brighter at the longer
wavelengths. The practical consequences of the contaminant sources are
thus:
\begin{itemize}
\item{The SMG in the Herschel beam boosts the 250\mic flux of the $z=0.35$
  galaxy; in cases near the threshold for detection this could push
  the $z=0.35$ source into the H-ATLAS catalogue. This effect results
  in an increased probability to find a nearby background SMG for
  250\mic sources close to the detection limit.}
\item{The lensing explanation for the over-density of SMG in these fields is not weakened by the
boosting effect, but the parameters derived from the modelling would
be affected. This is beyond the scope of this study.}
\item{The SMG in the beam reddens the {\em Herschel} sub-mm
  colours of the target galaxy, making them relatively brighter at 350
  and 500\mic than they would have been without the contaminant. This
  mimics the effect of colder dust in the SED and leads to an
  underestimate of the cold dust temperature in {\sc magphys} or
  two-component MBB fitting. For single MBB fits, it would result in a
  lower value for $\beta$, if that were a free parameter.}
\item{There are likely to be significant over-estimates of sub-mm
  fluxes if they are estimated by extrapolating SEDs fitted to {\em
    Herschel} fluxes which have been boosted in this way.}
\item{The combination of flux-boosting and reddening of the SED means
  that \Md and $\lsub^{\rm SED}$ can be biased high by 0.15-0.25 dex
  for samples affected by this process: 350--850\mic sources
  $0.2<z<0.7$ measured with large beams.}
\end{itemize}

\section{Conclusions}
We present serendipitous detections of high redshift dusty galaxies in
ALMA 850\mic images of a complete sample of twelve $z=0.35$ galaxies, selected at 250\mic from the H-ATLAS survey. 
\begin{itemize}
\item{Half of the ALMA Band 7 fields contained one or more high redshift SMG,
  an over-density of a factor 4--6 relative to the background counts.}
\item{We compared the statistics for the SMG with models of the
  lensing effect of group scale halos finding a remarkably good
  agreement, both in terms of the cross-correlation signal and the
  correlation between galaxies in halos with interacting satellites
  and the presence of SMG. Thus lensing is certainly a plausible
  explanation for the excess SMG detected around these sources.}
\item{These extra SMG contribute significantly to the SPIRE 350 and
  500\mic in some cases. We derive average flux-boosting factors of
  1.13, 1.26 and 1.44 for the 250, 350 and 500\mic SPIRE bands for
  this group of 12 representative 250\mic sources at $z=0.35$ from
  H-ATLAS. These are significantly larger corrections at 350 and
  500\mic than estimated by \citet{Valiante2016} who used simulations
  which did not include lensing between low and high-redshift
  sources. A boosting correction related to lensing is likely to be
  dependent on the redshifts of the target sources, as the probability
  for lensing depends on the lens-source geometry. For $z<0.1$ the
  lensing probability is much lower, but from $z=0.2-0.7$ it is likely
  that this sample is reasonably representative. Larger samples at
  different redshifts would be required to investigate this further.}
\end{itemize}

\section*{Data Availability}
The data underlying this article are publicly available from the ALMA archive
http://almascience.eso.org/aq/ using the project code 2012.1.00973.S. 

\section*{Acknowledgements} We thank S. Eales for helpful discussions on the impact of boosting on the H-ATLAS photometry, R. J. Ivison for healthy paranoia and an encyclopedic knowledge of SMG, and H. Gomez for careful consideration of the draft manuscript. LD and SJM acknowledge support from the European Research Council Advanced Investigator grant, COSMICISM and Consolidator grant, COSMIC DUST. LB and JGN acknowledge financial support from the PGC 2018 project PGC2018-101948-B-I00 (MICINN, FEDER) and PAPI-19-EMERG-11 (Universidad de Oviedo). JGN acknowledges financial support from the Spanish MINECO for the `Ramon y Cajal' fellowship (RYC-2013-13256).
This paper makes use of the following ALMA data:
ADS/JAO.ALMA\#2012.1.00973.S. ALMA is a partnership of ESO
(representing its member states), NSF (USA) and NINS (Japan), together
with NRC (Canada), MOST and ASIAA (Taiwan), and KASI (Republic of
Korea), in cooperation with the Republic of Chile. The Joint ALMA
Observatory is operated by ESO, AUI/NRAO and NAOJ. The National Radio
Astronomy Observatory is a facility of the National Science Foundation
operated under cooperative agreement by Associated Universities, Inc.
The {\it {\em Herschel}}-ATLAS is a project with {\it {\em Herschel}},
which is an ESA space observatory with science instruments provided by
European-led Principal Investigator consortia and with important
participation from NASA. The H-ATLAS web-site is
http://www.h-atlas.org. GAMA is a joint European-Australasian project
based around a spectroscopic campaign using the Anglo- Australian
Telescope. The GAMA input catalogue is based on data taken from the
Sloan Digital Sky Survey and the UKIRT Infrared Deep Sky
Survey. Complementary imaging of the GAMA regions is being obtained by
a number of independent survey programs including GALEX MIS, VST KIDS,
VISTA VIKING, WISE, {\em Herschel}-ATLAS, GMRT and ASKAP providing UV
to radio coverage. GAMA is funded by the STFC (UK), the ARC
(Australia), the AAO, and the participating institutions. The GAMA
website is: http://www.gama-survey.org/. This publication has made use
of data from the VIKING survey from VISTA at the ESO Paranal
Observatory, programme ID 179.A-2004. Data processing has been
contributed by the VISTA Data Flow System at CASU, Cambridge and WFAU,
Edinburgh.

\bibliographystyle{mnras/mnras}
\bibliography{Masterbib}

\bsp

\appendix

\section{Size fitting procedure}
\label{sec:inclination}
For the $u,v$ source fitting we use the {\sc CASA}  task
{\sc uvmodelfit} with a 2-d Gaussian model. The model is
described by 6 parameters: the flux $S_{\rm fit}$, position in RA and DEC,
major axis size $\theta_{m1}$, P.A., and the axial ratio $b/a$. For
our data, the axial ratio is very poorly constrained and this led to a
bias towards an axial ratio very close to zero, usually aligned with
the P.A. of the beam. This tendency produces a bias in the value for
$S_{\rm  fit}$ and $\theta_{m1}$, both of them being pushed towards the
maximum allowable values. Since an axial ratio approaching zero is not
physical, we conclude that no information can be derived on the axial
ratio of these sources. To avoid the bias we chose to constrain the
axial ratio to physically sensible values.

If we assume that galaxies are randomly oriented disks with
inclination $i$, the distribution of $\sin i$ will be uniform.  The
resulting distribution of the apparent axial ratios $(b/a)$ is
non-uniform such that:
$$b/a = \sqrt{\cos^2 i (1 - q^2) + q^2} $$ where $q$ is the intrinsic
axial ratio of the disk, which is thought to be between 0.1--0.2 for
disk galaxies.  For a randomly oriented distribution of disks the
expectation value of $\sin i$ is 0.5, which gives $\langle b/a \rangle
= 0.753$. We therefore fitted the model to the data with the axial ratio fixed
at $b/a = 0.753$ and these parameters are listed in Table~\ref{SMGT}.

The range of possible unknown disk orientations will contribute to the uncertainties of
the other parameters, even though we have fixed the axial ratio in the
fit. The upper and lower 1-$\sigma$ ranges of $b/a$ from the
distribution of disk orientations are represented by the 16th and 84th
percentiles ($b/a = 0.549, 0.987$) respectively.  So to estimate the
total uncertainties, we fit the model with $b/a$ fixed to its 16th and
84th percentile values, and use the resulting parameters to set the
1-$\sigma$ upper and lower bounds on $\theta_{m1}$ and $S_{\rm fit}$.
We then add this error in quadrature to the quoted fitting errors, $\delta x(\rm fit)$ to
arrive at our best estimate of the uncertainties:
$$\frac{\delta\theta_{m1}(\rm tot)}{\theta_{m1}} =
\sqrt{ \frac{\delta\theta_{m1}(\rm fit)}{\theta_{m1}}^2 + \left( \frac{\theta_{m1}(84)-\theta_{m1}(16)}{2\,\theta_{m1}}\right)^2}
$$
with a similar formalism for the other two parameters, $S_{\rm fit}$ and
$\theta_{m2}$ (which is simply the axial ratio times the major axis).

In the majority of cases, the variation of $b/a$ from the 16th--84th
percentile range causes little extra variation in $\theta_{m1}$ or
$S_{\rm fit}$, and so the quoted parameter is rather insensitive to the
exact value of axial ratio (1--3 percent in flux and 10--20 percent for
major axis). For two sources (SDP.4104, SDP.5347) there is a larger
sensitivity (5--7 percent in flux and 23--62 percent in major axis),
this is reflected in larger errors on the parameters.

\section{Notes on individual fields}
\label{sourcesS}

\subsubsection*{SDP.1160}
There are two bright SMG in this field located 5 and 12.5\asec from
the optical galaxy (Fig.~\ref{SMGF}); neither have a counterpart in
the VIKING K-band image ($\rm{K_{AB}>21.32}$). There is also a 500\mic
{\em Herschel} source 28.5\asec to the South that is identified in the
H-ATLAS `red source' catalogue, which uses a subtractive method to
recover sources which are very faint at 250\mic but brighter at
500\mic \citep{Ivison2016}. The H-ATLAS red source contributes 4.5~mJy
to the 500\mic flux for SDP.1160. This source is outside the Band 7
imaged area, but searching in the Band 3 cube we find tentative
evidence for two 3-mm line sources at similar redshift at the position
of the 350 and 500\mic peaks, see Table~\ref{candidateT}. Combining
the line information with the SPIRE colours (using the red source
catalogue fluxes) we find that the most plausible redshift is
$z=4.4-4.8$ with the lines being either CO(4--3) or \CIfull. A very
bright 850\mic source with a flux of 8--12~mJy would be expected in
this scenario, or a number of weaker sources which together produce
the red {\em Herschel} colours, such as those found by \citet{oteo18}.



\subsubsection*{SDP.2173}
There is an SMG in this field located 7.5\asec from the optical
galaxy, which has a K-band counterpart ($K_{AB}=21.45$) clearly
visible in the VIKING K-band image (Fig.~\ref{SMGF} and
Table~\ref{SMGT}). No 3-mm continuum or lines are detected at the location of the SMG.

\subsubsection*{SDP.3366}
The $z=0.35$ galaxy is interacting with a smaller companion to the
North-East, as evidenced from the \CI kinematics. Slightly further
north we find strong 850\mic continuum emission from a SMG coincident
with a very red source ($\rm{K_{AB}=19.0}$) located 3.4\asec to the
north-east of the $z=0.35$ galaxy (Fig~\ref{SMGF}). There is a second, fainter,
candidate SMG ($4-5\sigma$) just to the left of first, which is listed
in Table~\ref{candidateT} as 3366.s2. No 3-mm continuum or lines are
detected at the location of the SMG.

\subsubsection*{SDP.4104}
The $z=0.35$ galaxy is interacting with a similar sized companion, as
evidenced by the CO and \CI kinematics. There is a bright SMG in this
field located 7.2\asec from the largest member of the pair, which has
$\rm{K_{AB}=20.7}$ (see Fig.~\ref{SMGF}). No 3~mm continuum or lines
are detected at the location of the SMG.

\subsubsection*{SDP.5347}
The $z=0.35$ galaxy is interacting with several small satellites,
leaving a tidal trail of CO, \CI and some dust to the western side of
the galaxy. There is a 850\mic continuum source to the
north-east, which we assume to be a high redshift SMG; details are
presented in Table~\ref{SMGT} and Fig.~\ref{SMGF}. There is a faint whiff of K-band emission 
in a highly smoothed image, and no 3~mm continuum or line detection.

\subsubsection*{SDP.6418}
The $z=0.35$ galaxy is interacting with a smaller neighbour, as
evidenced by the \CI kinematics. The SMG in this image is strong
enough for self-calibration. There is no K-band counterpart and no
3~mm or line emission detected.

\subsubsection*{SDP.6451}
The SPIRE colours of this source are most definitely contaminated, as
is obvious when attempting an SED fit. We searched the data-set for
any evidence of sources which may be contributing to the red SPIRE
colours and find two potential candidates, which are listed in
Table~\ref{candidateT}. 6451.s1 is a 4.9$\sigma$ 850\mic source, which
is only just below the threshold we consider for calculation of the
number counts. 6451$^l$ refers to a potential 3-mm line source, which
just overlaps the B7 fov giving an upper limit to the 850\mic
flux. These are not considered in the over-density calculation but
listed in the Appendix for further information.

\section{Candidate high redshift sources.}
\begin{table*}
\caption{\label{candidateT} Properties of candidate SMG which are between 4-5$\sigma$ in continuum, or possible line detections in band 3.}
\begin{tabular}{lccccccc}
\hline
SMG  & R.A. & Dec. & $S_{850}$  & SNR & $r$  & $\rm{K_{AB}}$ & \\
     &      &      &  (mJy)    &     & (\asec)  &          &           \\ 
\hline
3366.s2   & 09:04:50.31 & $-$00:12:00.17 & $0.33 \pm 0.082^g$   &  5.1  & point & 4.2  & $<21.32$\\
          &             &                & $0.32 \pm 0.067^{pp}$ &  4.8  &       &     &         \\
6451.s1 & 09:08:50.0 & $+$02:26:00.85 & $0.68 \pm 0.14$ & 4.9 & 8.7 & $<21.32$& \\
\hline
SMG  &   R.A.       &  Dec.         & $S_{\rm line}$ & $\sigma_{\rm line}$ & $r$  & $\rm{K_{AB}}$  &  $\nu_{\rm line}$ \\
     &              &               &  (Jy\kms)    & (Jy\kms)          & (\asec) &            &   (GHz)    \\
\hline
1160.l1   & 09:00:29.32 & $+$01:21:46.8  & 2.83 & 0.76      & 30     & $<21.32$ & 85.25 \\
1160.l2   & 09:00:28.86 & $+$01:21:45.1  & 0.62 & 0.12       & 30     & $<21.32$ & 85.485 \\
1160.l3   & 09:00:29.24 & $+$01:21:46.8  & 0.56 & 0.12      & 30     & $<21.32$ & 87.525\\
\hline
5347.l1  & 09:06:59.22 & $+$02:02:20.74  & 1.73 & 0.52     & 26.9   & $<21.32$ & 85.948    \\
5347.l2  & 09:06:59.14 & $+$02:02:27.17  & 0.52 & 0.17     & 20.7   & $<21.32$ & 85.948    \\
\hline
6451.l1 & 09:08:48.6 & $+$02:26:00.8     & 1.64 & 0.37     &  13 & $<21.32$ & 99.1678\\
6451.l2 & 09:08:49.3 & $+$02:25:59.3     & 1.95 & 0.57     &  4     & $<21.32$ & 99.509\\ 
\hline 
\end{tabular}
\flushleft{\small{Line candidates at 3-mm. There is no continuum
    imaging at 850\mic at the location of the SDP.1160 line
    candidates. Plausible line identifications are: $z=3.06$ CO(3--2),
    $z=4.4/4.77$ CO(4--3)/\CIfull and $z=5.76$ CO(5--4), but only the
    $z=4.5-4.8$ solutions fit well with the SPIRE colours. }}
\end{table*}

\label{lastpage}
\end{document}